\documentclass[journal]{IEEEtran}

\usepackage{cite} 

\usepackage[caption=false,font=footnotesize]{subfig}	
\usepackage{graphicx}

\usepackage{algorithm}
\usepackage{algorithmicx}
\usepackage{algpseudocode}

\usepackage{amsmath,amsfonts,amssymb,amsthm}
\usepackage{mathtools}
\usepackage{mathrsfs}	
\usepackage{subdepth}	
\usepackage{nicefrac}

\usepackage{array}	
\usepackage{tabularx}   

\usepackage{hyperref}
\usepackage[sort,compress,capitalise]{cleveref}	
\usepackage{url}

\usepackage[dvipsnames]{xcolor} 
\usepackage{xspace} 

\usepackage{tikz}
\usetikzlibrary{babel}  
\usepackage[straightvoltages,nooldvoltagedirection]{circuitikz}

{
\theoremstyle{plain}
\newtheorem{Hypothesis}{Hypothesis}

}

{
\theoremstyle{definition}

}

\crefname{Hypothesis}{Hyp.}{Hyps.}
\Crefname{Hypothesis}{Hyp.}{Hyps.}

\crefname{Lemma}{Lemma}{Lemmata}
\Crefname{Lemma}{Lemma}{Lemmata}

\crefname{Definition}{Def.}{Defs.}
\Crefname{Definition}{Def.}{Defs.}

\newcommand{\CIGRE}{CIGR{\'E}\xspace}

\newcommand{\DFT}{\textup{DFT}\xspace}  

\newcommand{\RMS}{RMS\xspace}   

\newcommand{\LTI}{LTI\xspace}	
\newcommand{\LTP}{LTP\xspace}	

\newcommand{\AC}{AC\xspace} 
\newcommand{\DC}{DC\xspace}	

\newcommand{\CIDER}[1][]{CIDER#1\xspace}    
\newcommand{\TE}[1][]{\text{TE#1}\xspace}	

\newcommand{\TDS}{\textup{TDS}\xspace}	
\newcommand{\HPF}{\textup{HPF}\xspace}	

\newcommand{\EMTP}{EMTP\xspace}		

\newcommand{\PLL}[1][]{PLL#1\xspace}	
\newcommand{\PWM}[1][]{PWM#1\xspace}	

\newcommand{\KPI}[1][]{KPI#1\xspace}    

\newcommand{\RealNum}{\mathbb{R}}   

\newcommand{\Abs}[1]{\left|#1\right|}

\newcommand{\Cos}[1]{\cos\left(#1\right)}
\newcommand{\Sin}[1]{\sin\left(#1\right)}
\newcommand{\Exp}[1]{\exp\left(#1\right)}

\newcommand{\Set}[1]{\mathcal{#1}}

\newcommand{\diag}{\operatorname{diag}}

\newcommand{\XT}{\mathbf{x}}
\newcommand{\UT}{\mathbf{u}}
\newcommand{\WT}{\mathbf{w}}
\newcommand{\YT}{\mathbf{y}}

\newcommand{\AP}{\mathbf{A}}
\newcommand{\BP}{\mathbf{B}}
\newcommand{\CP}{\mathbf{C}}
\newcommand{\DP}{\mathbf{D}}
\newcommand{\EP}{\mathbf{E}}
\newcommand{\FP}{\mathbf{F}}

\newcommand{\grid}{\gamma}
\newcommand{\pwr}{\pi}
\newcommand{\ctrl}{\kappa}
\newcommand{\act}{\alpha}
\newcommand{\fltr}{\varphi}
\newcommand{\trafo}{\tau}
\newcommand{\refr}{\rho}
\newcommand{\spt}{\sigma}

\newcommand{\Y}{\mathbf{Y}}     

\newcommand{\phases}{\Set{P}}

\newcommand{\phsABC}{\textup{ABC}\xspace}

\newcommand{\cmpD}{\textup{D}\xspace}
\newcommand{\cmpQ}{\textup{Q}\xspace}

\newcommand{\cmpDQ}{\textup{DQ}\xspace}

\newcommand{\harmonics}{\Set{H}}

\newcommand{\VT}{\mathbf{v}}    
\newcommand{\IT}{\mathbf{i}}    

\newcommand{\RP}{\mathbf{R}}
\newcommand{\LP}{\mathbf{L}}
\newcommand{\GP}{\mathbf{G}}
\newcommand{\KP}{\mathbf{K}}

\newcommand{\ctrlP}{\textup{P}\xspace} 

\newcommand{\ctrlPI}{\textup{PI}\xspace}
\newcommand{\ctrlPID}{\textup{PID}\xspace}
\newcommand{\ctrlPR}{\textup{PR}\xspace}

\newcommand{\ctrlFB}{\textup{FB}\xspace}   
\newcommand{\ctrlFF}{\textup{FF}\xspace}   
\newcommand{\ctrlFT}{\textup{FT}\xspace}   

\newcommand{\FBoT}[1]{\frac{\KP_{\ctrlFB,#1}}{T_{\ctrlFB,#1}}}	
\newcommand{\FFpFB}[1]{\KP_{\ctrlFF,#1}+\KP_{\ctrlFB,#1}}   	
\newcommand{\eye}[1]{\diag(\mathbf{1}_{#1})}                    
\newcommand{\zero}[2]{\mathbf{0}_{#1\times#2}}                    

\newcommand{\TP}{\mathbf{T}}

\allowdisplaybreaks

\begin{document}



\title{\huge{
    Harmonic Power-Flow Study of Polyphase Grids\\
	with Converter-Interfaced Distributed Energy Resources,\\
	Part II: Model Library and Validation
}}



\author{%
	Johanna~Kristin~Maria~Becker,~\IEEEmembership{Member,~IEEE},
	Andreas~Martin~Kettner,~\IEEEmembership{Member,~IEEE},
	Lorenzo~Reyes-Chamorro,~\IEEEmembership{Senior~Member,~IEEE},
	Zhixiang~Zou,~\IEEEmembership{Member,~IEEE},\\
	Marco~Liserre,~\IEEEmembership{Fellow,~IEEE},
	and~Mario~Paolone,~\IEEEmembership{Senior~Member,~IEEE}%
	\thanks{J. Becker, and M. Paolone are with the Distributed Electrical Systems Laboratory at the {\'E}cole Polytechnique F{\'e}d{\'e}rale de Lausanne (EPFL) in CH-1015 Lausanne, Switzerland (E-mail: \{johanna.becker, mario.paolone\}@epfl.ch).}%
	\thanks{A. Kettner is with PSI NEPLAN AG, 8700 Küsnacht, Switzerland (E-mail: andreas.kettner@neplan.ch).}%
	\thanks{L. Reyes-Chamorro is with the Facultad de Ciencias de la Ingenier{\'i}a at the Universidad Austral de Chile (UACh) in CL-5111187 Valdivia, Chile (E-mail: lorenzo.reyes@uach.cl).}%
	\thanks{Z. Zou is with the School of Electrical Engineering, Southeast University, in PRC-210096 Nanjing, China (E-mail: zzou@seu.edu.cn).}%
	\thanks{M. Liserre is with the Chair of Power Electronics at the Christian-Albrechts-Universit{\"a}t zu Kiel (CAU) in DE-24143 Kiel, Germany (E-mail: ml@tf.uni-kiel.de).}%
	\thanks{This work was funded by the Schweizerischer Nationalfonds (SNF, Swiss National Science Foundation) via the National Research Programme NRP~70 ``Energy Turnaround'' (NRP 70 "Energy Turnaround" (projects nr. 173661 and 197060) and by the Deutsche Forschungsgemeinschaft (DFG, German Research Foundation) via the Priority Programme DFG~SPP~1984 ``Hybrid and Multimodal Energy Systems'' (project nr. 359982322).}%
}

\maketitle







\begin{abstract}
    In Part~I, a method for the \emph{Harmonic Power-Flow} (\HPF) study of three-phase power grids with \emph{Converter-Interfaced Distributed Energy Resources} (\CIDER[s]) is proposed.
    The method is based on generic and modular representations of the grid and the \CIDER[s], and explicitly accounts for coupling between harmonics.
    In Part~II, the \HPF method is validated.
    First, the applicability of the modeling framework is demonstrated on typical grid-forming and grid-following \CIDER[s].
    Then, the \HPF method is implemented in Matlab and compared against time-domain simulations with Simulink.
    The accuracy of the models and the performance of the solution algorithm are assessed for individual resources and a modified version of the \CIGRE low-voltage benchmark microgrid (i.e., with additional unbalanced components).
    The observed maximum errors are 6.3E-5~p.u. w.r.t. voltage magnitude, 1.3E-3~p.u. w.r.t. current magnitude, and 0.9~deg w.r.t. phase.
    Moreover, the scalability of the method is assessed w.r.t. the number of \CIDER[s] and the maximum harmonic order ($\leqslant$25).
    For the maximum problem size, the execution time of the \HPF method is 6.52~sec, which is 5 times faster than the time-domain simulation.
    The convergence of the method is robust w.r.t. the choice of the initial point, and multiplicity of solutions has not been observed.
\end{abstract}



\begin{IEEEkeywords}
	Distributed energy resources,
	harmonic power-flow study,
	polyphase power systems,
	power electronic converters,
	unbalanced power grids.
\end{IEEEkeywords}



\section*{Nomenclature}

\begin{center}
    
\begin{tabularx}{\columnwidth}{p{3cm}p{5cm}}
    \hline
    \multicolumn{2}{c}{Component Models}\\
    \hline
\end{tabularx}
\begin{IEEEdescription}[\IEEEusemathlabelsep\IEEEsetlabelwidth{$(\cdot)_{+/-/0}$}]
    \item[$\pwr$]
        The power hardware of a \CIDER
    \item[$\ctrl$]
        The control software of a \CIDER
    \item[$\act$]
        The actuator of a \CIDER
    \item[$\lambda$]
        A stage in the cascaded structure of a \CIDER ($\lambda\in\{1,\ldots,\Lambda\}$)
    \item[$\VT_{\lambda}$]
        The voltage in a capacitive filter stage $\lambda$
    \item[$\IT_{\lambda}$]
        The current in an inductive filter stage $\lambda$
    \item[$\RP_{\lambda},\LP_{\lambda}$]
        The compound electrical parameters of an inductive filter stage $\lambda$
    \item[$\GP_{\lambda},\CP_{\lambda}$]
        The compound electrical parameters of a capacitive filter stage $\lambda$
    \item[$\KP_{\ctrlFB,\lambda}$]
        The feed-back gain of a controller stage $\lambda$
    \item[$T_{\ctrlFB,\lambda}$]
        The integration time of a controller stage $\lambda$ 
    \item[$\KP_{\ctrlFF,\lambda}$]
        The feed-forward gain of a controller stage $\lambda$
    \item[$\KP_{\ctrlFT,\lambda}$]
        The feed-through gain of a controller stage $\lambda$
\end{IEEEdescription}
\hrule
\end{center}

\begin{center}
    
\begin{tabularx}{\columnwidth}{p{3cm}p{5cm}}
    \hline
    \multicolumn{2}{c}{Resource Models}\\
    \hline
\end{tabularx}
\begin{IEEEdescription}[\IEEEusemathlabelsep\IEEEsetlabelwidth{$(\cdot)_{+/-/0}$}]
    \item[$\grid$]
        The power grid
    \item[$\refr$] 
        The reference calculation of a \CIDER
    \item[$\spt$] 
        The setpoint of a \CIDER
    \item[$f_1$]
        The fundamental frequency ($f_{1}\coloneqq\frac{1}{T}$)
    \item[$h\in\harmonics$] 
        A harmonic order ($\harmonics\coloneqq\{-h_{\max},\ldots,h_{\max}\}$)
    \item[$\theta$]
        A given reference angle
    \item[$\trafo_{\ctrl|\pwr}$]
        A change of reference frame from $\pwr$ to $\ctrl$
    \item[$\mathbf{T}_{\ctrl|\pwr}(t)$] 
        The \LTP matrix which describes $\trafo_{\ctrl|\pwr}$
    \item[$\mathbf{T}_{\ctrl|\pwr,h}$]
        The Fourier coefficient of $\TP_{\pwr|\grid}$ ($h\in\harmonics$)
    \item[$\hat{\mathbf{T}}_{\ctrl|\pwr}$]
        The Toeplitz matrix composed of the $\hat{\mathbf{T}}_{\ctrl|\pwr,h}$
    \item[$\mathbf{x}(t)$]
        The state vector of a state-space model
    \item[$\mathbf{u}(t)$]
        The input vector of a state-space model
    \item[$\mathbf{y}(t)$]
        The output vector of a state-space model
    \item[$\mathbf{w}(t)$]
        The disturbance vector of a state-space model
    \item[$\mathbf{A}(t)$]
        The system matrix of an \LTP system
    \item[$\mathbf{B}(t)$]
        The input matrix of an \LTP system
    \item[$\mathbf{C}(t)$]
        The output matrix of an \LTP system
    \item[$\mathbf{D}(t)$]
        The feed-through matrix of an \LTP system
    \item[$\mathbf{E}(t)$]
        The input disturbance matrix of an \LTP system
    \item[$\mathbf{F}(t)$]
        The output disturbance matrix of an \LTP system
    \item[$V_\spt$]
        The voltage setpoint of a grid-forming resource
    \item[$S_\spt$]
        The power setpoint of a grid-following resource
    \item[$v_{\grid,\cmpD}(t)$]
        The direct component of $v_{\grid}$
    \item[$v_{\grid,\cmpQ}(t)$]
        The quadrature component of $v_{\grid}$
    \item[$\xi_\cmpD(t)$]
        The time-variant signal content of $v_{\grid,\cmpD}(t)$
    \item[$V_{\grid,\cmpD,h}$]
        The Fourier coefficient of $v_{\grid,\cmpD}(t)$ ($h\in\harmonics$)
\end{IEEEdescription}
\hrule
\end{center}

\begin{center}
    
\begin{tabularx}{\columnwidth}{p{3cm}p{5cm}}
    \hline
    \multicolumn{2}{c}{Validation}\\
    \hline
\end{tabularx}
\begin{IEEEdescription}[\IEEEusemathlabelsep\IEEEsetlabelwidth{$(\cdot)_{+/-/0}$}]
    \item[$\mathbf{X}_{h}$]
        The Fourier coefficient of a polyphase electrical quantity ($h\in\harmonics$)
    \item[$e_{\textup{abs}}(\mathbf{X}_{h})$]
        The maximum absolute error over all phases w.r.t. $\Abs{\mathbf{X}_{h}}$ between \HPF and \TDS
    \item[$e_{\textup{arg}}(\mathbf{X}_{h})$]
        The maximum absolute error over all phases w.r.t. $\angle{\mathbf{X}_{h}}$ between \HPF and \TDS
    \item[$(\cdot)_{+/-/0}$]
        Positive-, negative-, and homopolar-sequence components of a three-phase electrical quantity.
    \item[$T_{exc,\HPF}$]
        The execution time of the \HPF
    \item[$\mu(\cdot)$]
        The mean value operator
    \item[$\sigma(\cdot)$]
        The standard deviation operator
\end{IEEEdescription}
\hrule

\end{center}

\section{Introduction}
\label{sec:intro}



\IEEEPARstart{T}{he} validation of the proposed \emph{Harmonic Power-Flow} (\HPF) method involves two aspects.
Firstly, the suitability of the modeling framework to represent common types of \emph{Converter-Interfaced Distributed Energy Resources} (\CIDER[s]) with grid-forming or grid-following controls has to be confirmed.
In this respect, power converters equipped with either LC or LCL filters are considered.
Secondly, the accuracy of the modeling framework and the performance of the solution algorithm need to be assessed.
To this end, the proposed method is applied to analyze individual resources as well as an entire system.
More precisely, the \HPF algorithm is implemented in Matlab and compared against time-domain simulations carried out in Simulink.



The remainder of Part~II is organized as follows:
First, the models of standard components of \CIDER[s] (i.e., actuator, filters, and controllers) are developed in \cref{sec:lib-cmp}, including a thorough discussion of their properties and working principles.
These component models are combined in \cref{sec:lib-rsc} to construct the complete models of a grid-forming and a grid-following \CIDER.
Then, the proposed method is tested on individual resources in \cref{sec:val-rsc}, as well as on the \CIGRE low-voltage benchmark microgrid in \cref{sec:val-sys}.
Finally, the conclusions are drawn in \cref{sec:concl}.

\section{Library of Component Models}
\label{sec:lib-cmp}

In this section, the models of actuators, filters, and controllers are developed.
Note that, in order to obtain compact formulas, the time-dependency of the electrical quantities and signals is not explicitly stated each time.


\subsection{Actuator}
\label{sec:lib-cmp:act}

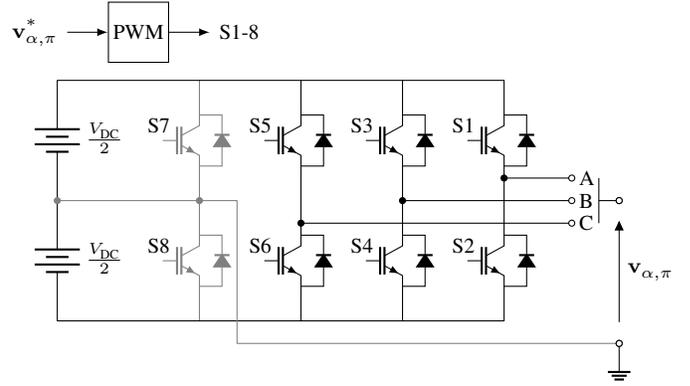
\begin{figure}
    \centering
    {

\footnotesize
\ctikzset{bipoles/length=1.0cm}
\ctikzset{bipoles/diode/height=.2}
\ctikzset{bipoles/diode/width=.15}
\tikzstyle{block}=[rectangle, draw=black, minimum size=8mm, inner sep=0pt]

\begin{circuitikz}
    
	\def\x{1.8}
	\def\y{1.6}
	\def\dlabel{-0.3*\x}
	
	
    \coordinate (SP) at ($(0,\y)$);
    \coordinate (SC) at ($(0,0)$);
    \coordinate (SN) at ($(0,-\y)$);
    \coordinate (LNP) at ($(1.05*\x,\y)$);
    \coordinate (LNN) at ($(1.05*\x,-\y)$);
    \coordinate (LNC) at ($(1.05*\x,0)$);
    \coordinate (LAP) at ($(LNP)+0.75*(\x,0)$);
    \coordinate (LAN) at ($(LNN)+0.75*(\x,0)$);
    \coordinate (LAC) at ($(LNC)+0.75*(\x,0)$);
    \coordinate (LBP) at ($(LAP)+0.75*(\x,0)$);
    \coordinate (LBN) at ($(LAN)+0.75*(\x,0)$);
    \coordinate (LBC) at ($(LAC)+0.75*(\x,0)$);
    \coordinate (LCP) at ($(LBP)+0.75*(\x,0)$);
    \coordinate (LCN) at ($(LBN)+0.75*(\x,0)$);
    \coordinate (LCC) at ($(LBC)+0.75*(\x,0)$);

    \node[nigbt,scale=0.8,label={[shift={(\dlabel,0)}]S5}] (DAp) at ($(LAC)+(0,0.5*\y)$) {};
    \draw (DAp.E)++(0,0.1) -- ++(0.3,0) to[D*] ($(DAp.C)+(0.3,-0.1)$)   -- ++(-0.3,0);
    \node[nigbt,scale=0.8,label={[shift={(\dlabel,0)}]S6}] (DAn) at ($(LAC)-(0,0.5*\y)$) {};
    \draw (DAn.E)++(0,0.1) -- ++(0.3,0) to[D*] ($(DAn.C)+(0.3,-0.1)$)   -- ++(-0.3,0);
    \node[nigbt,scale=0.8,label={[shift={(\dlabel,0)}]S3}] (DBp) at ($(LBC)+(0,0.5*\y)$) {};
    \draw (DBp.E)++(0,0.1) -- ++(0.3,0) to[D*] ($(DBp.C)+(0.3,-0.1)$)   -- ++(-0.3,0);
    \node[nigbt,scale=0.8,label={[shift={(\dlabel,0)}]S4}] (DBn) at ($(LBC)-(0,0.5*\y)$) {};
    \draw (DBn.E)++(0,0.1) -- ++(0.3,0) to[D*] ($(DBn.C)+(0.3,-0.1)$)   -- ++(-0.3,0);
    \node[nigbt,scale=0.8,label={[shift={(\dlabel,0)}]S1}] (DCp) at ($(LCC)+(0,0.5*\y)$) {};
    \draw (DCp.E)++(0,0.1) -- ++(0.3,0) to[D*] ($(DCp.C)+(0.3,-0.1)$)   -- ++(-0.3,0);
    \node[nigbt,scale=0.8,label={[shift={(\dlabel,0)}]S2}] (DCn) at ($(LCC)-(0,0.5*\y)$) {};
    \draw (DCn.E)++(0,0.1) -- ++(0.3,0) to[D*] ($(DCn.C)+(0.3,-0.1)$)   -- ++(-0.3,0);
    
    \draw (DAp.C) to[short] (LAP)
          (LAP) to[short] (LCP)
          (LCP) to[short] (DCp.C);
    \draw (DBp.C) to[short] (LBP);
    
    \draw (DAn.E) to[short] (LAN)
          (LAN) to[short] (LCN)
          (LCN) to[short] (DCn.E);
    \draw (DBn.E) to[short] (LBN);
    
    \draw (DAp.E) to[short] (DAn.C);
    \draw (DBp.E) to[short] (DBn.C);
    \draw (DCp.E) to[short] (DCn.C);
    
    \node[nigbt,scale=0.8,label={[shift={(\dlabel,0)}]S7},color=gray] (DNp) at ($(LNC)+(0,0.5*\y)$) {};
    \draw[gray] (DNp.E)++(0,0.1) -- ++(0.3,0) to[D*,color=gray] ($(DNp.C)+(0.3,-0.1)$)   -- ++(-0.3,0);
    \node[nigbt,scale=0.8,label={[shift={(\dlabel,0)}]S8},color=gray] (DNn) at ($(LNC)-(0,0.5*\y)$) {};
    \draw[gray] (DNn.E)++(0,0.1) -- ++(0.3,0) to[D*,color=gray] ($(DNn.C)+(0.3,-0.1)$)   -- ++(-0.3,0);
    
    \draw[gray] (DNp.E) to[short] (DNn.C);   
    \draw[gray] (DNn.E) to[short] (LNN); 
    \draw[gray] (DNp.C) to[short] (LNP);

    \draw (LAP) to[short] (SP)
          (SP)  to[battery, l=$\frac{V_{\textup{DC}}}{2}$] (SC)
          (SC)  to[battery, l=$\frac{V_{\textup{DC}}}{2}$] (SN)
          (SN)  to[short] (LAN);

    \coordinate (TA) at ($(LAC)+(0,0.3) + 2*0.75*(\x,0)+ 0.5*(\x,0)$);
    \coordinate (TB) at ($(LBC) + 1*0.75*(\x,0)+ 0.5*(\x,0)$);
    \coordinate (TC) at ($(LCC)-(0,0.3) + 0.5*(\x,0)$);
    \coordinate (TN) at ($(LCN)-(0,0.3) + 0.5*(\x,0)$);
	\coordinate (G) at ($(TN) + 0.35*(\x,0)$);
    
    \draw ($(LCC)+(0,0.3)$) to[short,*-o] (TA) node[right] {A};
    \draw (LBC) to[short,*-o] (TB) node[right] {B};
    \draw ($(LAC)-(0,0.3)$) to[short,*-o] (TC) node[right] {C};
    \draw[gray] (SC) to[short,*-,color=gray] (LNC)
                (LNC) to[short,*-,color=gray] ($(LNC)+(0.5,0)$)
                ($(LNC)+(0.5,0)$) to[short] ($(LNN)+(0.5,-0.3)$)
                ($(LNN)+(0.5,-0.3)$) to[short,-*] (G);
    
	
	\coordinate (P) at ($(0.6*\x,1.4*\y)$);
	
	\node[block] (PWM) at (P) {PWM};
	\draw[-latex] (PWM.east) to node[right,align=left]{$~~~$S1-8} ($(PWM.east) + 0.3*(\x,0)$);
	\draw[-latex] ($(PWM.west)-0.3*(\x,0)$) to node[left,align=left]{$\VT^{*}_{\act,\pwr}~~$} (PWM.west);
	

	\draw ($(G)-(0,0.05*\y)$) node[ground](GND){} to (G);
	
	\draw ($(TA) + 0.2*(\x,0)$) to[short] ($(TC) + 0.2*(\x,0)$);
	\draw ($(TB) + 0.2*(\x,0)$) to[short,-o] ($(TB) + 0.35*(\x,0)$);
	
	\draw (G) to[open,o-o,v=${\VT_{\act,\pwr}}$] ($(TB) + 0.35*(\x,0)$);
	
\end{circuitikz}
}
    
    \caption
    {%
        Schematic diagram of a three-phase two-level power converter, which is commonly used for \CIDER[s].
        The fourth leg is optional: it is required only if the power converter has to be able to inject or absorb homopolar currents.
    }
    \label{fig:act}
\end{figure}

The actuator is the power converter which interfaces the \DC side of the \CIDER (i.e., a source or load) with its \AC side (i.e., the filter).
It consists of an array of switches (i.e., power-transistor-type devices), which are controlled so that the output voltage of the actuator $\VT_{\act}$ follows the reference $\VT^{*}_{\act}$ (see \cref{fig:act}).
The switching signals are typically generated by a \emph{Pulse-Width Modulator} (\PWM).
This modulation creates distortions in the output voltage of the power converter.
Indeed, this is why a filter is needed.
Typically, these distortions occur at high frequencies (i.e., several kHz), which are far beyond the frequency range that is of interest for harmonic analysis (i.e., up to 1-2 kHz) \cite{Std:BSI-EN-50160:2000}.
Hence, the actuator can be regarded as ideal voltage source in the \HPF study:
\begin{Hypothesis}\label{hyp:cmp:act}
    In the frequency range of interest for the \HPF study, switching effects are negligible.
    Therefore, the actuator is regarded as ideal voltage source:
    \begin{alignat}{2}
        \VT_{\act,\pwr} = \VT^{*}_{\act,\pwr}&~\in~   \RealNum^{\dim(\pwr)\times1}
    \end{alignat}
\end{Hypothesis}
\noindent
Note that the size of this vector depends on the reference frame in which the power hardware is modelled%
\footnote{%
    If phase coordinates are used, $\VT_{\act,\pwr},\VT^{*}_{\act,\pwr}\in\RealNum^{3\times1}$.
}.%

\subsection{Filter Stages}
\label{sec:lib-cmp:fltr}

\begin{figure}[t]
	\centering
	
	\subfloat[]
	{%
		\centering
		{

\footnotesize
\ctikzset{bipoles/length=1.0cm}

\begin{circuitikz}
	
	\def\x{1.6}
	\def\y{2.0}
	
	\coordinate (NL) at (-\x,0);
	\coordinate (PL) at ($(NL)+(0,\y)$);	
	\coordinate (NR) at (\x,0);
	\coordinate (PR) at ($(NR)+(0,\y)$);
	
	\draw (NL) to[short] (NR);
	\draw (PL)
		to[short] ($7/8*(PL)+1/8*(PR)$)
		to[resistor=$\mathbf{R}_{\lambda,\pwr}$] ($1/2*(PL)+1/2*(PR)$)
		to[inductor=$\mathbf{L}_{\lambda,\pwr}$,v=${\mathbf{i}_{\lambda,\pwr}}$] ($1/8*(PL)+7/8*(PR)$)
		to[short] (PR);
		
	\draw (NL) to[open,v=${\mathbf{v}_{\lambda-1,\pwr}}$,o-o] (PL);
	\draw (NR) to[open,v^=${\mathbf{v}_{\lambda+1,\pwr}}$,o-o] (PR);
	
\end{circuitikz}

}
		\label{fig:fltr:ind}
	}
	\subfloat[]
	{%
		\centering
		{

\footnotesize
\ctikzset{bipoles/length=1.0cm}

\begin{circuitikz}
	
	\def\x{2.0}
	\def\y{2.0}
	
	\coordinate (NL) at (-\x,0);
	\coordinate (PL) at ($(NL)+(0,\y)$);	
	\coordinate (NR) at (\x,0);
	\coordinate (PR) at ($(NR)+(0,\y)$);
	
	\draw (NL) to[short,o-o] (NR);
	
	\draw (PL)
		to[short,i=${\mathbf{i}_{\lambda-1,\pwr}}$,o-] ($2/3*(PL)+1/3*(PR)$)
		to[short] ($1/3*(PL)+2/3*(PR)$)
		to[short,i=${\mathbf{i}_{\lambda+1,\pwr}}$,-o] (PR);
	
	\draw ($2/3*(NL)+1/3*(NR)$) to[resistor=$\mathbf{G}_{\lambda,\pwr}$,*-*] ($2/3*(PL)+1/3*(PR)$);
	\draw ($1/3*(NL)+2/3*(NR)$) to[capacitor=$\mathbf{C}_{\lambda,\pwr}$,v=${\mathbf{v}_{\lambda,\pwr}}$,*-*] ($1/3*(PL)+2/3*(PR)$);
		
\end{circuitikz}

}
		\label{fig:fltr:cap}
	}
	
	\caption
	{%
		Equivalent circuits of a filter stage $\lambda$ constructed from inductors (\ref{fig:fltr:ind}) or capacitors (\ref{fig:fltr:cap}), respectively.
		Observe that voltages, currents, and electrical parameters are expressed in the reference frame of the power hardware $\pwr$.
	}
	\label{fig:fltr}
\end{figure}
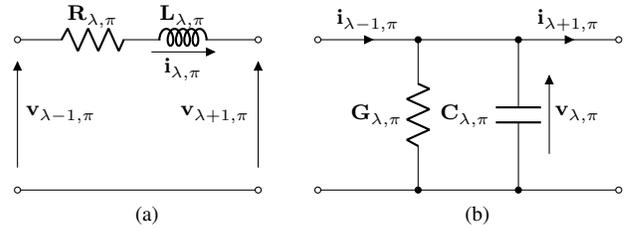

In order to attenuate the high-frequency distortions resulting from the switching in the actuator, \CIDER[s] are equipped with cascades of filter stages \cite{Jrn:PSE:PEC:2004:Blaabjerg}.
Each stage consists of inductors or capacitors, which filter currents or voltages, respectively.
Notably, the commonly used L-, LC-, and LCL-filters as well as higher-order filters are built in this way.

Consider any stage $\lambda$ in the cascade of filters.
An inductive filter stage is represented by the equivalent circuit in \cref{fig:fltr:ind}, which is described by the following differential equation:
\begin{equation}
		\VT_{\lambda-1,\pwr} - \VT_{\lambda+1,\pwr}
	=	\RP_{\lambda,\pwr}\IT_{\lambda,\pwr} + \LP_{\lambda,\pwr}\frac{d}{dt}\IT_{\lambda,\pwr}
	\label{eq:fltr:ind:diffeq}
\end{equation}
$\RP_{\lambda,\pwr},\LP_{\lambda,\pwr}\in\RealNum^{\dim(\pwr)\times\dim(\pwr)}$ are the compound electrical parameters of the inductive filter stage, $\IT_{\lambda,\pwr}\in\RealNum^{\dim(\pwr)\times1}$ is the current flowing through it, and $\VT_{\lambda-1,\pwr},\VT_{\lambda+1,\pwr}\in\RealNum^{\dim(\pwr)\times1}$ are the voltages at the start and end node of the stage, respectively.
Again, the sizes of these vectors and matrices depend on the reference frame in which the power hardware is modelled.
A capacitive filter stage is represented by the equivalent circuit in \cref{fig:fltr:cap}, which is described by
\begin{equation}
		\IT_{\lambda-1,\pwr} - \IT_{\lambda+1,\pwr}
	=	\GP_{\lambda,\pwr}\VT_{\lambda,\pwr} + \CP_{\lambda,\pwr}\frac{d}{dt}\VT_{\lambda,\pwr}
	\label{eq:fltr:cap:diffeq}
\end{equation}
$\GP_{\lambda,\pwr},\CP_{\lambda,\pwr}\in\RealNum^{\dim(\pwr)\times\dim(\pwr)}$ are the compound electrical parameters of the capacitive filter stage, $\VT_{\lambda,\pwr}\in\RealNum^{\dim(\pwr)\times1}$ is the voltage across it, and $\IT_{\lambda-1,\pwr},\IT_{\lambda+1,\pwr}\in\RealNum^{\dim(\pwr)\times1}$ are the currents flowing into and out of the stage, respectively.

In practice, the filter stages are built from identical discrete elements (i.e., one element per phase).
Accordingly, the following hypothesis can be made:
\begin{Hypothesis}\label{hyp:cmp:fltr}
    The compound electrical parameters of the filter stages are diagonal matrices with equal nonzero entries.
    That is, an inductive filter stage is characterized by
    \begin{equation}
        \RP_{\lambda,\pwr} = R_{\lambda}\diag(\mathbf{1}_{\pwr}),~
        \LP_{\lambda,\pwr} = L_{\lambda}\diag(\mathbf{1}_{\pwr})
        \label{eq:fltr:ind:param}
    \end{equation}
    and a capacitive filter stage by
    \begin{equation}
        \GP_{\lambda,\pwr} = G_{\lambda}\diag(\mathbf{1}_{\pwr}),~
        \CP_{\lambda,\pwr} = C_{\lambda}\diag(\mathbf{1}_{\pwr})
        \label{eq:fltr:cap:param}
    \end{equation}
    where $\diag(\mathbf{1}_{\pwr})$ is the identity matrix w.r.t. reference frame of the power hardware.
    $R_{\lambda}$, $L_{\lambda}$ and $G_{\lambda}$, $C_{\lambda}$ are the parameters of the discrete elements.
\end{Hypothesis}


\subsection{Controller Stages}
\label{sec:lib-cmp:ctrl}

\begin{figure}[t]
	\centering
	\hspace{-0.5cm}
	\subfloat[]
	{%
		\centering
		{

\footnotesize

\tikzstyle{block}=[rectangle, draw=black, minimum size=5.5mm, inner sep=0pt]
\tikzstyle{sum}=[circle, draw=black, minimum size=2mm, inner sep=0pt]
\tikzstyle{dot}=[circle, draw=black, fill=black, minimum size=1mm, inner sep=0pt]
\tikzstyle{signal}=[-latex]

\begin{tikzpicture}
	
	
	\def\x{0.85}
	\def\y{0.85}
	
	
	
	\coordinate (RI) at (0,0); 
	
	
	
	\node[dot] (FFI) at ($(RI)-(3/5*\x,0)$) {};
	\node[block] (FFC) at ($(FFI)-(\x,0)$) {\ctrlFF};	
	\node[sum] (FFO) at ($(FFC)-(\x,0)$) {$+$};
	
	\draw (RI.west) to node[below]{$\IT^{*}_{\lambda,\ctrl}$} (FFI.east);
	\draw[-latex] (FFI.west) to (FFC.east);
	\draw[signal] (FFC.west) to (FFO.east);
	
	
	
	\node[sum] (FBI) at ($(FFI)+(0,\y)$) {$+$};
	\coordinate (FBM) at ($(FBI)+(0,\y)$);
	\node[block] (FBC) at ($(FFC)+(0,\y)$) {\ctrlFB};
	\coordinate (FBO) at ($(FFO)+(0,\y)$);
	
	\draw[signal] (FFI) to (FBI.south);
	\draw[signal] (FBM) to node[at start,above]{$\IT_{\lambda,\ctrl}$} node[at end,right]{$-$} (FBI.north);
	\draw[signal] (FBI.west) to (FBC.east);
	\draw[signal] (FBC.west) to (FBO) to (FFO.north);
	
	
	
	\node[sum] (FTO) at ($(FFO)-(4/5*\x,0)$) {$+$};
	\node[block] (FTC) at ($(FTO)+(0,\y)$) {\ctrlFT};
	\coordinate (FTM) at ($(FTC)+(0,\y)$);
	
	\draw[signal] (FFO.west) to (FTO.east);
	\draw[signal] (FTM) to node[at start,above]{$\VT_{\lambda+1,\ctrl}$} (FTC.north);
	\draw[signal] (FTC.south) to (FTO.north);
	
	
	
	\coordinate (RO) at ($(FTO)-(4/5*\x,0)$);
	
	\draw[signal] (FTO.west) to node[below]{$\VT^{*}_{\lambda-1,\ctrl}$} (RO.east);
	
\end{tikzpicture}

}
		\label{fig:ctrl:ind}
	}
	\hspace{-0.8cm}
	\subfloat[]
	{%
		\centering
		{

\footnotesize

\tikzstyle{block}=[rectangle, draw=black, minimum size=5.5mm, inner sep=0pt]
\tikzstyle{sum}=[circle, draw=black, minimum size=2mm, inner sep=0pt]
\tikzstyle{dot}=[circle, draw=black, fill=black, minimum size=1mm, inner sep=0pt]
\tikzstyle{signal}=[-latex]

\begin{tikzpicture}
	
	
	\def\x{0.85}
	\def\y{0.85}
	
	
	
	\coordinate (RI) at (0,0);
	
	
	
	\node[dot] (FFI) at ($(RI)-(3/5*\x,0)$) {};
	\node[block] (FFC) at ($(FFI)-(\x,0)$) {\ctrlFF};	
	\node[sum] (FFO) at ($(FFC)-(\x,0)$) {$+$};
	
	\draw (RI.west) to node[below]{$\VT^{*}_{\lambda,\ctrl}$} (FFI.east);
	\draw[-latex] (FFI.west) to (FFC.east);
	\draw[signal] (FFC.west) to (FFO.east);
	
	
	
	\node[sum] (FBI) at ($(FFI)+(0,\y)$) {$+$};
	\coordinate (FBM) at ($(FBI)+(0,\y)$);
	\node[block] (FBC) at ($(FFC)+(0,\y)$) {\ctrlFB};
	\coordinate (FBO) at ($(FFO)+(0,\y)$);
	
	\draw[signal] (FFI) to (FBI.south);
	\draw[signal] (FBM) to node[at start,above]{$\VT_{\lambda,\ctrl}$} node[at end,right]{$-$} (FBI.north);
	\draw[signal] (FBI.west) to (FBC.east);
	\draw[signal] (FBC.west) to (FBO) to (FFO.north);
	
	
	
	\node[sum] (FTO) at ($(FFO)-(4/5*\x,0)$) {$+$};
	\node[block] (FTC) at ($(FTO)+(0,\y)$) {\ctrlFT};
	\coordinate (FTM) at ($(FTC)+(0,\y)$);
	
	\draw[signal] (FFO.west) to (FTO.east);
	\draw[signal] (FTM) to node[at start,above]{$\IT_{\lambda+1,\ctrl}$} (FTC.north);
	\draw[signal] (FTC.south) to (FTO.north);
	
	

	\coordinate (RO) at ($(FTO)-(4/5*\x,0)$);
	
	\draw[signal] (FTO.west) to node[below]{$\IT^{*}_{\lambda-1,\ctrl}$} (RO.east);
	
\end{tikzpicture}

}
		\label{fig:ctrl:cap}
	}
	
	\caption
	{%
		Block diagrams of a controller stage $\lambda$ associated with an inductive (\ref{fig:ctrl:ind}) and capacitive (\ref{fig:ctrl:cap}) filter stage, respectively.
		In general, the control law includes \emph{Feed-Back} (\ctrlFB), \emph{Feed-Forward} (\ctrlFF), and \emph{Feed-Through} (\ctrlFT) terms.
		Observe that voltages and currents are expressed in the reference frame of the control software $\ctrl$.
	}
	\label{fig:ctrl}
\end{figure}
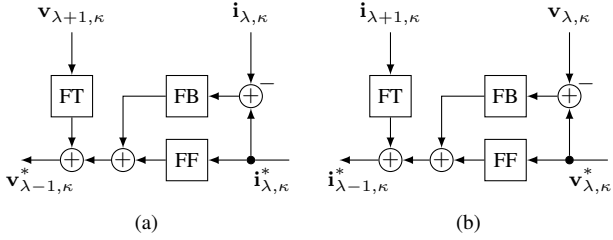
Each filer stage can be coupled with a corresponding controller, which regulates either the current through or the voltage across the filter element, depending on whether the filter stage is inductive or capacitive.
\footnote{%
    As already mentioned in Part~I of this paper, it is common practice for \CIDER[s] with LCL filter to implement one instead of two current control loops.
    Notably, this can easily be represented in the proposed framework.
}
As illustrated in \cref{fig:ctrl}, such a controller generally performs \emph{Feed-Back} (\ctrlFB), \emph{Feed-Forward} (\ctrlFF), and \emph{Feed-Through} (\ctrlFT) control.
More precisely, as shown in \cref{fig:ctrl:ind,fig:ctrl:cap}, a stage $\lambda$ of the controller calculates the reference for the next-inner stage $\lambda-1$ from the deviation between its own state and the desired reference (i.e., via \ctrlFB and \ctrlFF control), as well as the state of the next-outer stage $\lambda+1$ (i.e., via \ctrlFT control).
In principle, each block of a controller stage could be composed of multiple subblocks connected in series or in parallel (e.g., for the mitigation of specific harmonics).
In practice, simple \emph{Proportional-Integral-Derivative} (\ctrlPID) and \emph{Proportional-Resonant} (\ctrlPR) controllers are commonly used \cite{Jrn:Blaabjerg:2006}.


For the sake of illustration, \ctrlPI-controllers are considered for \ctrlFB control, and \ctrlP-controllers for \ctrlFF and \ctrlFT control:
\begin{Hypothesis}\label{hyp:cmp:ctrl}
    Each controller stage consists of a \ctrlPI controller for \ctrlFB control, and two \ctrlP controllers for \ctrlFF and \ctrlFT controls.
\end{Hypothesis}
\noindent
Let $\KP_{\ctrlFB,\lambda}$, $\KP_{\ctrlFF,\lambda},\KP_{\ctrlFT,\lambda}\in\RealNum^{\dim(\ctrl)\times\dim(\ctrl)}$ be the proportional gains and $T_{\ctrlFB,\lambda}$ the integration time, respectively.
The control law for an inductive filter stage is given by (see \cref{fig:ctrl:ind})
\begin{align}
		\VT^{*}_{\lambda-1,\ctrl}
	&=	\left[~
			\begin{aligned}
				    &\KP_{\ctrlFB,\lambda}\left(\Delta\IT_{\lambda,\ctrl}+\frac{1}{T_{\ctrlFB,\lambda}}\int\Delta\IT_{\lambda,\ctrl}\,dt\right)\\
			    +   &\KP_{\ctrlFT,\lambda}\VT_{\lambda+1,\ctrl} + \KP_{\ctrlFF,\lambda}\IT^{*}_{\lambda,\ctrl}
			\end{aligned}
			\right.
	\label{eq:ctrl:ind:law}\\
				\Delta\IT_{\lambda,\ctrl}
	&\coloneqq	\IT^{*}_{\lambda,\ctrl}-\IT_{\lambda,\ctrl}
	\label{eq:ctrl:ind:error}
\end{align}
$\VT^{*}_{\lambda-1,\ctrl},\IT^{*}_{\lambda,\ctrl}\in\RealNum^{\dim(\ctrl)\times1}$ are the reference voltage at the input of the controller stage and the reference current at its output, respectively.
$\VT_{\lambda+1,\ctrl},\IT_{\lambda,\ctrl}\in\RealNum^{\dim(\ctrl)\times1}$ are the voltage at the output of the filter stage and the current through it, respectively, expressed in the reference frame of the controller.
The control law for a capacitive filter stage is analogous (see \cref{fig:ctrl:cap}):
\begin{align}
		\IT^{*}_{\lambda-1,\ctrl}
	&=	\left[~
		\begin{aligned}
			&\KP_{\ctrlFB,\lambda}\left(\Delta\VT_{\lambda,\ctrl}+\frac{1}{T_{\ctrlFB,\lambda}}\int\Delta\VT_{\lambda,\ctrl}\,dt\right)\\
		+	&\KP_{\ctrlFT,\lambda}\IT_{\lambda+1,\ctrl} + \KP_{\ctrlFF,\lambda}\VT^{*}_{\lambda,\ctrl}
		\end{aligned}
		\right.
	\label{eq:ctrl:cap:law}
	\\
	            \Delta\VT_{\lambda,\ctrl}
	&\coloneqq	\VT^{*}_{\lambda,\ctrl}-\VT_{\lambda,\ctrl}
	\label{eq:ctrl:cap:error}
\end{align}
$\IT^{*}_{\lambda-1,\ctrl},\VT^{*}_{\lambda,\ctrl}\in\RealNum^{\dim(\ctrl)\times1}$ are the reference current at the input of the controller stage and the reference voltage at its output, respectively.
$\IT_{\lambda+1,\ctrl},\VT_{\lambda,\ctrl}\in\RealNum^{\dim(\ctrl)\times1}$ are the current at the output of the filter stage and voltage across it, respectively, expressed in the reference frame of the controller.


Typically, the \ctrlFB and \ctrlFT controllers treat each coordinate in the reference frame separately, and apply equal gains to them.
In line with this fact, the following hypothesis is made:
\begin{Hypothesis}\label{hyp:ctrl:fb-ft}
	The \ctrlFB and \ctrlFT gains are diagonal matrices with equal nonzero entries.
	That is
	\begin{align}
				\KP_{\ctrlFB,\lambda}
		&=	K_{\ctrlFB,\lambda}\diag(\mathbf{1}_{\ctrl})
		\\
				\KP_{\ctrlFT,\lambda}
		&=	K_{\ctrlFT,\lambda}\diag(\mathbf{1}_{\ctrl})
	\end{align}
	where $\diag(\mathbf{1}_{\ctrl})$ is the identity matrix w.r.t. the reference frame of the controls software.
\end{Hypothesis}
The \ctrlFF controllers, by contrast, are obtained by restating the dynamical models of the filter stages, which are given in the reference frame of the power hardware, in the reference frame of the control software \cite{Jrn:Wang:2015}.
To this end, the transformation matrices $\TP_{\ctrl|\pwr}\in\RealNum^{\dim(\ctrl)\times\dim(\pwr)}$ and $\TP_{\pwr|\ctrl}\in\RealNum^{\dim(\pwr)\times\dim(\ctrl)}$ are substituted into the filter equations \eqref{eq:fltr:ind:diffeq}--\eqref{eq:fltr:cap:diffeq}.
For the inductive filter stage, one obtains
\begin{equation}
        \VT_{\lambda-1,\ctrl} - \VT_{\lambda+1,\ctrl}
	=   \RP_{\lambda,\ctrl}\IT_{\lambda,\ctrl} + \LP_{\lambda,\ctrl}\frac{d}{dt}\IT_{\lambda,\ctrl}
	\label{eq:fltr:ind:diffeq:ctrl}
\end{equation}
where $\RP_{\lambda,\ctrl},\LP_{\lambda,\ctrl}\in\RealNum^{\dim(\ctrl)\times\dim(\ctrl)}$ are given by
\begin{align}
        \RP_{\lambda,\ctrl}
    &=      \TP_{\ctrl|\pwr}\RP_{\lambda,\pwr}\TP_{\pwr|\ctrl}
        +   \TP_{\ctrl|\pwr}\LP_{\lambda,\pwr}\frac{d}{dt}\TP_{\pwr|\ctrl}
    \label{eq:fltr:ind:rst:ctrl}\\
        \LP_{\lambda,\ctrl}
    &=  \TP_{\ctrl|\pwr}\LP_{\lambda,\pwr}\TP_{\pwr|\ctrl}
\end{align}
Analogously, for the capacitive filter stage, one finds
\begin{equation}
        \IT_{\lambda-1,\ctrl} - \IT_{\lambda+1,\ctrl}
	=	\GP_{\lambda,\ctrl}\VT_{\lambda,\ctrl} + \CP_{\lambda,\ctrl}\frac{d}{dt}\VT_{\lambda,\ctrl}
	\label{eq:fltr:cap:diffeq:ctrl}
\end{equation}
where $\GP_{\lambda,\ctrl},\CP_{\lambda,\ctrl}\in\RealNum^{\dim(\ctrl)\times\dim(\ctrl)}$ are given by
\begin{align}
        \GP_{\lambda,\ctrl}
    &=      \TP_{\ctrl|\pwr}\GP_{\lambda,\pwr}\TP_{\pwr|\ctrl}
        +   \TP_{\ctrl|\pwr}\CP_{\lambda,\pwr}\frac{d}{dt}\TP_{\pwr|\ctrl}
    \label{eq:fltr:cap:cnd:ctrl}\\
        \CP_{\lambda,\ctrl}
    &=  \TP_{\ctrl|\pwr}\CP_{\lambda,\pwr}\TP_{\pwr|\ctrl}
\end{align}
Observe that the expressions for $\RP_{\lambda,\ctrl}$ and $\GP_{\lambda,\ctrl}$ include terms which result from the temporal derivatives of $\IT_{\lambda,\pwr}=\TP_{\pwr|\ctrl}\IT_{\lambda,\ctrl}$ and $\VT_{\lambda,\pwr}=\TP_{\pwr|\ctrl}\VT_{\lambda,\ctrl}$, respectively.
These expressions often turn out to be time-invariant thanks to \cref{hyp:cmp:fltr}.
For instance, as will be shown shortly, this is the case in the $\cmpDQ$ frame.

Using \eqref{eq:fltr:ind:diffeq:ctrl} and \eqref{eq:fltr:cap:diffeq:ctrl}, the \ctrlFF gains can be set in order to achieve zero error in steady-state (e.g., \cite{Jrn:Wang:2015}) via the following additional hypothesis.
\begin{Hypothesis}\label{hyp:cmp:ctrl:ff}
    The \ctrlFF gains are set to
    \begin{equation}
            \KP_{\ctrlFF,\lambda}
        =   \left\{
            \begin{array}{cl}
                    \RP_{\lambda,\ctrl}
                &   \text{for inductive filter stages}\\
                    \GP_{\lambda,\ctrl}
                &   \text{for capacitive filter stages}
            \end{array}
            \right.
    \end{equation}
\end{Hypothesis}

\section{Library of Resource Models}
\label{sec:lib-rsc}

In this section, the models of typical grid-forming and grid-following \CIDER[s] are constructed from the components presented in \cref{sec:lib-cmp}, based on the the generic \LTP models defined in Part~I.


\subsection{Circuit Configurations and Reference Frames}
\label{sec:lib-rsc:frames}



Usually, the grid and the power hardware are both modeled in phase coordinates, but their circuit configuration may not be the same.
The grid is a four-wire system (i.e., three phase plus one neutral conductor), whereas the power converters can be either four-leg or three-leg devices (i.e., with or without neutral conductor).
If a \CIDER with a four-leg power converter is connected to a four-wire grid, all sequence components (i.e., positive, negative, and homopolar sequences) of voltage and current can pass in both directions.
This corresponds to
\begin{equation}
    \TP_{\pwr|\grid} = \TP_{\grid|\pwr} = \diag(\mathbf{1}_{3})
    \label{eq:trafo:grid:neutral}
\end{equation}
By contrast, if a \CIDER with a three-leg power converter is connected to a four-wire grid, homopolar sequences are blocked in both directions.
This is represented by
\begin{equation}
    \TP_{\pwr|\grid} = \TP_{\grid|\pwr} = \diag(\mathbf{1}_{3}) - \frac{1}{3} \mathbf{1}_{3\times3}
    \label{eq:trafo:grid:no_neutral}
\end{equation}



It is common practice to implement the control software in \emph{Direct-Quadrature} ($\cmpDQ$) components \cite{Jrn:Blaabjerg:2006}.
If the power hardware is modeled in phase ($\phsABC$) coordinates, as previously mentioned, one obtains
\begin{align}
        \mathbf{T}_{\ctrl|\pwr}
	&=	\mathbf{T}_{\pwr|\ctrl}^{T}
	\label{eq:trafo:pwr-to-ctrl}
	\\
		\mathbf{T}_{\pwr|\ctrl}
	&=	\sqrt{\frac{2}{3}}
		\begin{bmatrix}
				\Cos{\theta}
			&	-\Sin{\theta}
			\\
				\Cos{\theta-\frac{2\pi}{3}}
			&	-\Sin{\theta-\frac{2\pi}{3}}
			\\
				\Cos{\theta+\frac{2\pi}{3}}
			&	-\Sin{\theta+\frac{2\pi}{3}}
		\end{bmatrix}
	\label{eq:trafo:ctrl-to-pwr}
\end{align}
where $\theta$ is a given reference angle.
How the reference angle is obtained, depends on the type of \CIDER.
Namely, it is calculated from a reference clock in grid-forming \CIDER[s], whereas a synchronisation unit is needed for grid-following \CIDER[s].
In either case, it is assumed that $\theta$ is synchronized with the fundamental tone:
\begin{Hypothesis}\label{hyp:src:frame}
    Irrespective of the type of \CIDER, the reference angle $\theta$, w.r.t. which the $\cmpDQ$ frame is defined, is given by
    \begin{equation}
        \theta = 2\pi f_{1}t + \theta_{0}
        \label{eq:trafo:angle}
    \end{equation}
    where $\theta_{0}$ is a known offset.
\end{Hypothesis}
\noindent
If this hypothesis holds, the Fourier coefficients of the transformation matrices \eqref{eq:trafo:pwr-to-ctrl}--\eqref{eq:trafo:ctrl-to-pwr}, which are needed for the harmonic-domain model, are given as follows:
\begin{align}
        \mathbf{T}_{\pwr|\ctrl,+1}
    &=  \sqrt{\frac{2}{3}} \Exp{j \theta_0}
    \begin{bmatrix}
			\frac{1}{2}
		&	-\frac{1}{2j}
		\\
			\frac{1}{2} \alpha^{*}
		&	-\frac{1}{2j} \alpha^{*}
		\\
			\frac{1}{2} \alpha
		&	-\frac{1}{2j} \alpha
    \end{bmatrix}\\
        \mathbf{T}_{\pwr|\ctrl,-1}
    &=  \mathbf{T}_{\pwr|\ctrl,+1}^{*}
\end{align}
where $\alpha = \Exp{j \frac{2\pi}{3}}$.
As explained in Part~I, the Fourier coefficients of time-periodic matrices appear on the diagonals of the associated Toeplitz matrices in the harmonic domain.
For example, the coefficients of order $h=\pm1$ appear on the first upper and lower diagonal, respectively.
Accordingly, $\Hat{\mathbf{T}}_{\ctrl|\pwr}$ and $\Hat{\mathbf{T}}_{\pwr|\ctrl}$ have a block band structure, which introduces coupling between the harmonics.



Having specified the reference frames, the \ctrlFF gain $\KP_{\ctrlFF,\lambda}$ as given in \cref{hyp:cmp:ctrl:ff} can be evaluated.
By substitution of \eqref{eq:trafo:pwr-to-ctrl}--\eqref{eq:trafo:ctrl-to-pwr} and \cref{hyp:src:frame} into \eqref{eq:fltr:ind:rst:ctrl} and \eqref{eq:fltr:cap:cnd:ctrl}, one can show that
\begin{align}
            \RP_{\lambda,\cmpDQ}
    	=	\begin{bmatrix}
				R_\lambda		    &-2\pi f_1L_\lambda\\
				2\pi f_1 L_\lambda	&R_\lambda
			\end{bmatrix}\\
			\GP_{\lambda,\cmpDQ}
		=	\begin{bmatrix}
				G_\lambda	        &-2\pi f_ 1 C_\lambda\\
				2\pi f_1 C_\lambda	&G_\lambda	
			\end{bmatrix}
\end{align}
The off-diagonal elements are a.k.a. decoupling terms \cite{Jrn:Rocabert:2012}.


\subsection{Grid-Forming Resource}
\label{sec:lib-rsc:forming}

\begin{figure}[t]
	\centering
	{

\ctikzset{bipoles/length=1.0cm}
\tikzstyle{block}=[rectangle, draw=black, minimum size=8mm, inner sep=0pt]
\tikzstyle{dot}=[circle, draw=black, fill=black, minimum size=2pt, inner sep=0pt]
\tikzstyle{measurement}=[rectangle,draw=black,minimum size=1mm,inner sep=0pt]
\tikzstyle{signal}=[-latex]

\def\transform#1#2
{%
\begin{scope}[shift={#2}]
    \node[block] (#1) at (0,0) {};
    \draw (#1.south west) to (#1.north east);
    \node at ($(#1.north west)+(0.3,-0.15)$) {\scriptsize$\phsABC$};
    \node at ($(#1.south east)+(-0.2,0.15)$) {\scriptsize$\cmpDQ$};
\end{scope}
}
    
\footnotesize

\begin{circuitikz}
    
    \def\x{1.6}
    \def\y{1.6}
    
    
    
    \coordinate (AN) at (0,0);
    \coordinate (AP) at ($(AN)+(0,\y)$);
    
    \coordinate (FALN) at ($(AN)+(0.5*\x,0)$);
    \coordinate (FALP) at ($(FALN)+(0,\y)$);
    \coordinate (FARN) at ($(FALN)+(\x,0)$);
    \coordinate (FARP) at ($(FARN)+(0,\y)$);
    
    \coordinate (FGLN) at ($(FARN)+(\x,0)$);
    \coordinate (FGLP) at ($(FGLN)+(0,\y)$);
    \coordinate (FGRN) at ($(FGLN)+(\x,0)$);
    \coordinate (FGRP) at ($(FGRN)+(0,\y)$);
    
    \draw (FALN)
        to[short] (AN)
        to[voltage source] (AP)
        to[short] (FALP);
    \draw (FALN) to[open,v_=$\VT_{\act,\phsABC}$] (FALP);
    
    \draw (FALN) to[short] (FARN);
    \draw (FALP) to[generic=${R_\act,L_\act}$] (FARP);
    
    \draw (FARN) to[short] (FGLN);
    \draw (FARP)
        to[short,i=$\IT_{\alpha,\phsABC}$] ($0.5*(FARP)+0.5*(FGLP)$)
        to[short] (FGLP); 
    
    \draw ($0.5*(FARN)+0.5*(FGLN)$) to[generic=${G_\fltr,C_\fltr}$,v=$\VT_{\fltr,\phsABC}$,*-*] ($0.5*(FARP)+0.5*(FGLP)$);
    
    \draw (FGLN) to[short,-o] (FGRN);
    \draw (FGLP) to[short,-o,i=$\IT_{\grid,\phsABC}$] (FGRP);
    
    
    
    \coordinate (IA) at ($(AN)-0.1*(0,\y)$);
    \coordinate (OFA) at ($(IA)+(\x,0)$);
    \coordinate (OFG) at ($(OFA)+(\x,0)$);
    \coordinate (OG) at ($(OFG)+(\x,0)$);
    
    \transform{TA}{($(IA)-0.8*(0,\y)$)}
    \transform{TFA}{($(TA)+(\x,0)$)}
    \transform{TFG}{($(TFA)+(\x,0)$)}
    \transform{TG}{($(TFG)+(\x,0)$)}
    
    \draw[signal] (TA.north) to node[midway,left]{$\VT_{\act,\phsABC}$} (IA.south);
    \draw[signal] (OFA) to node[midway,left]{$\IT_{\act,\phsABC}$} (TFA.north);
    \draw[signal] (OFG) to node[midway,left]{$\VT_{\fltr,\phsABC}$} (TFG.north);
    \draw[signal] (OG) to node[midway,left]{$\IT_{\grid,\phsABC}$} (TG.north);
    
    
    
    \node[block] (CFA) at ($(TFA)-1.4*(0,\y)$) {\ctrlPI};
    \node[block] (CFG) at ($(CFA)+(\x,0)$) {\ctrlPI};
    \node[block] (R) at ($(CFG)+(\x,0)$) {$\begin{bmatrix}*\\0\end{bmatrix}$};
    \node (S) at ($(R)+0.7*(\x,0)$) {$V_{\spt}$};
    
    \node[dot] (XFG) at ($0.5*(TFG)+0.5*(CFG)$) {};
    
    \draw[signal] (TG.south)
        to node[midway,left]{$\IT_{\grid,\cmpDQ}$} ($0.5*(TG)+0.5*(R)$)
        to ($0.5*(TFG.south)+0.5*(CFG.north east)$)
        to ($0.5*(CFG.north)+0.5*(CFG.north east)$);
        
    \draw[-] (TFG.south) to node[midway,left]{$\VT_{\fltr,\cmpDQ}$} (XFG.north);
    \draw[signal] (XFG.south) to (CFG.north);
    \draw[signal] (XFG.west)
        to  ($0.5*(TFA.south)+0.5*(CFA.north east)$)
        to ($0.5*(CFA.north)+0.5*(CFA.north east)$);
    
    \draw[signal] (TFA.south)
        to node[midway,left]{$\IT_{\act,\cmpDQ}$} ($0.5*(TFA.south)+0.5*(CFA.north)$)
        to (CFA.north);
    
    \draw[signal] (R.west) to node[midway,above]{$\VT^*_{\fltr,\cmpDQ}$} (CFG.east);
    \draw[signal] (CFG.west) to node[midway,above]{$\IT^*_{\act,\cmpDQ}$} (CFA.east);
    
    \draw[signal] (CFA.west)
        to ($(CFA)-(\x,0)$)
        to node[near end,left]{$\VT_{\act,\cmpDQ}$} (TA.south);
        
    \draw[signal] (S.west) to (R.east);
        
\end{circuitikz}

}
	\caption
	{%
		Schematic diagram of a grid-forming \CIDER with an LC filter.
	}
	\label{fig:CIDER:forming}
\end{figure}

\cref{fig:CIDER:forming} shows the schematic diagram of a typical grid-forming \CIDER.
Its power hardware consists of a \PWM actuator and an LC filter, and its control software of a two-stage \ctrlPI controller.
The actuator is a four-leg power converter which can inject or absorb homopolar currents.
This feature is of crucial importance for islanded operation, during which the grid-forming \CIDER takes the role of the slack.



The state of the power hardware is given by the inductor current $\IT_{\act,\phsABC}\in\RealNum^{3\times1}$ and the capacitor voltage $\VT_{\fltr,\phsABC}\in\RealNum^{3\times1}$.
The input and disturbance are the actuator voltage $\VT_{\act,\phsABC}\in\RealNum^{3\times1}$ and grid current $\IT_{\grid,\phsABC}\in\RealNum^{3\times1}$, respectively.
The output includes both state and disturbance.
That is
\begin{alignat}{2}
    \XT_\pwr(t)
	&=	\begin{bmatrix}
			\IT_{\act,\phsABC}(t)\\
			\VT_{\fltr,\phsABC}(t)
		\end{bmatrix}
    &~\in~\RealNum^{6\times1}
	\label{eq:vf:pwr:state}
	\\
	    \UT_\pwr(t)
	&=	\VT_{\act,\phsABC}(t)
    &~\in~\RealNum^{3\times1}
    \\
	    \WT_\pwr(t)
	&=	\IT_{\grid,\phsABC}(t)
    &~\in~\RealNum^{3\times1}
    \\
		\YT_\pwr(t)
	&=	\begin{bmatrix}
	        \XT_\pwr(t)\\
	        \WT_\pwr(t)
	   \end{bmatrix}
    &~\in~\RealNum^{9\times1}
	\label{eq:vf:pwr:output}
\end{alignat}
The time-domain state-space model of the power hardware is obtained by combining the differential equations \eqref{eq:fltr:ind:diffeq} and \eqref{eq:fltr:cap:diffeq} of the filter stages.
This yields
\begin{align}
		\AP_\pwr(t)
	&=  \begin{bmatrix}
			-\LP_\act^{-1}\RP_\act	&-\LP_\act^{-1}\\
			\CP_\fltr^{-1}			&-\CP_\fltr^{-1}\GP_\fltr
		\end{bmatrix}
	\\
		\BP_\pwr(t)
	&=	\begin{bmatrix}
	        \LP_\act^{-1}\\
	        \mathbf{0}_{3\times3}
	    \end{bmatrix}
	\\
	    \EP_{\pwr}(t)
	&=	\begin{bmatrix}
	        \mathbf{0}_{3\times3}\\
	        -\CP_\fltr^{-1}
	    \end{bmatrix}
	\\
        \CP_\pwr(t)
	&=	\begin{bmatrix}
	        \diag(\mathbf{1}_6)\\
	        \mathbf{0}_{3\times6}
	    \end{bmatrix}
	\\
	    \DP_{\pwr}(t)
	&=	\mathbf{0}_{9\times3}
	\\
	    \FP_{\pwr}(t)
	&=	\begin{bmatrix}
	        \mathbf{0}_{6\times3}\\
	        \diag(\mathbf{1}_3)
	    \end{bmatrix}
\end{align}
The sizes of these matrices follow directly from \eqref{eq:vf:pwr:state}--\eqref{eq:vf:pwr:output}.
Note that these matrices are time-invariant, which means that only the Fourier coefficient for $h=0$ is nonzero.
For instance:
\begin{equation}
    \AP_\pwr(t) = \AP_{\pwr,0}
\end{equation}
The same holds for the other matrices of the state-space model.
Accordingly, the power hardware is an \LTI system, which is a particular case of an \LTP system.



Recall from \cref{hyp:cmp:ctrl} that each controller stage consists of one \ctrlPI controller (i.e., for \ctrlFB control) and two \ctrlP controllers (i.e., for \ctrlFF and \ctrlFT control).
Since the control software is composed of \ctrlPI controllers, its state is given by the temporal integrals of the errors w.r.t. the inductor current $\Delta\IT_{\act,\cmpDQ}\in\RealNum^{2\times1}$ and the capacitor voltage $\Delta\VT_{\fltr,\cmpDQ}\in\RealNum^{2\times1}$.
Its input and output are defined by the interconnection with the power hardware as shown in \cref{fig:CIDER:forming}.
The disturbance is the reference voltage $\VT^*_{\fltr,\cmpDQ}\in\RealNum^{2\times1}$ of the outer controller stage.
Accordingly
\begin{alignat}{2}
	    \XT_\ctrl(t)
	&=	\int
		\begin{bmatrix}
			\Delta\IT_{\act,\cmpDQ}(t)	\\
			\Delta\VT_{\fltr,\cmpDQ}(t)
		\end{bmatrix}
		dt
    &~\in~\RealNum^{4\times1}
	\label{eq:vf:ctrl:state}
	\\
	    \UT_\ctrl(t)
	&=	\begin{bmatrix}
			\IT_{\act,\cmpDQ}(t)	\\
			\VT_{\fltr,\cmpDQ}(t)	\\
			\IT_{\grid,\cmpDQ}(t)
		\end{bmatrix}
    &~\in~\RealNum^{6\times1}
    \\
	    \WT_\ctrl(t)
	&=	\VT^*_{\fltr,\cmpDQ}(t)
    &~\in~\RealNum^{2\times1}
    \\
		\YT_\ctrl(t)
	&=	\VT_{\act,\cmpDQ} (t)
    &~\in~\RealNum^{2\times1}
	\label{eq:vf:ctrl:output}
\end{alignat}
The time-domain state-space model of the control software is found by combining the differential equations \eqref{eq:ctrl:ind:law} and \eqref{eq:ctrl:cap:law} of the controller stages.
This gives
\begin{align}
		\AP_\ctrl(t)
	&=  \begin{bmatrix}
	            \zero{2}{2}
	        &   \FBoT{\fltr}\\
	            \zero{2}{2}
	        &   \zero{2}{2}
	    \end{bmatrix}
	\\
		\BP_\ctrl(t)
	&=	\begin{bmatrix}
			-\eye{2}	&-\KP_{\ctrlFB,\fltr}   & \KP_{\ctrlFT,\fltr}\\
			\zero{2}{2}	&-\eye{2}   & \zero{2}{2}      
		\end{bmatrix}
	\\
	    \EP_{\ctrl}(t)
	&=	\begin{bmatrix}
	        \FFpFB{\fltr}\\
	        \eye{2}
	    \end{bmatrix}
	\\
        \CP_\ctrl(t)
	&=	\begin{bmatrix}
	        \FBoT{\act} &	(\FFpFB{\act})\FBoT{\fltr}  
	    \end{bmatrix}
	\\  
	    \DP_{\ctrl}(t)
	&=	\begin{bmatrix}
	        -\KP_{\ctrlFB,\act} & (\DP_{\ctrl})_2 & (\DP_{\ctrl})_3
	    \end{bmatrix}
    \\   
	(\DP_{\ctrl})_2
	&=  \KP_{\ctrlFT,\act}-(\FFpFB{\act})\KP_{\ctrlFB,\fltr}  
	\\   
	(\DP_{\ctrl})_3
	&=  (\FFpFB{\act})\KP_{\ctrlFT,\fltr}  
	\\
	    \FP_{\ctrl}(t)
	&=	(\FFpFB{\act})(\FFpFB{\fltr})
\end{align}
The sizes of these matrices follow directly from \eqref{eq:vf:ctrl:state}--\eqref{eq:vf:ctrl:output}.
Evidently, the control software is an \LTI system, too.
Indeed, this is one of the reasons for the popularity of the \cmpDQ frame since its invention almost a century ago \cite{Jrn:Park:1929}.



In grid-forming \CIDER[s], the reference angle $\theta$ is computed from the frequency setpoint $f_{\spt}$ through integration over time
\begin{equation}
    \theta = 2\pi\int f_{\spt}\,dt = 2\pi f_{\spt}t
\end{equation}
Hence, in line with \cref{hyp:src:frame}, the following hypothesis is made.
\begin{Hypothesis}\label{hyp:cmp:trafo:forming}
    In steady state, the frequency setpoints of all grid-forming \CIDER[s] are equal to the fundamental frequency:
    \begin{equation}
        f_{\spt}=f_{1}
    \end{equation}
\end{Hypothesis}
\noindent
Indeed, if the grid-forming \CIDER[s] attempted to impose incompatible frequencies on the power system, no steady-state equilibrium could exist.
As discussed in Part~I, the setpoints for the resource-level controllers are determined by system-level controllers.
Since system-level controllers act on a substantially slower timescale than resource-level ones, the steady-state values of the setpoints can be determined in an independent analysis (i.e., prior to the \HPF study).

The reference voltage $\VT^*_{\fltr,\cmpDQ}$ is calculated from the voltage setpoint $V_\spt$ as follows:
\begin{Hypothesis}\label{hyp:rsc:former:ref}
    The reference voltage for the grid-forming \CIDER[s] is calculated as follows
    \begin{equation}
            \VT^*_{\fltr,\cmpDQ}(t)
        =   \sqrt{\frac{3}{2}}
            \begin{bmatrix}
                V_\spt\\
                0
            \end{bmatrix}
    \end{equation}
    where $V_\spt$ is the setpoint for the peak voltage.
\end{Hypothesis}


\subsection{Grid-Following Resource}
\label{sec:lib-rsc:following}

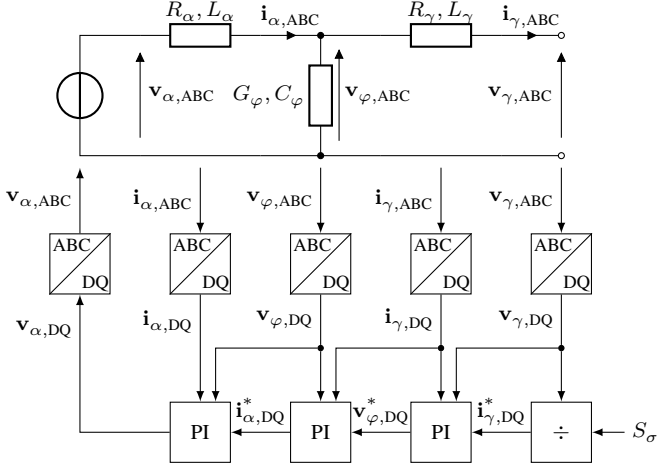
\begin{figure}[t]
	\centering
	   {

\ctikzset{bipoles/length=1.0cm}
\tikzstyle{block}=[rectangle, draw=black, minimum size=8mm, inner sep=0pt]
\tikzstyle{dot}=[circle, draw=black, fill=black, minimum size=2pt, inner sep=0pt]
\tikzstyle{measurement}=[rectangle,draw=black,minimum size=1mm,inner sep=0pt]
\tikzstyle{signal}=[-latex]

\def\transform#1#2
{%
\begin{scope}[shift={#2}]
    \node[block] (#1) at (0,0) {};
    \draw (#1.south west) to (#1.north east);
    \node at ($(#1.north west)+(0.3,-0.15)$) {\scriptsize$\phsABC$};
    \node at ($(#1.south east)+(-0.2,0.15)$) {\scriptsize$\cmpDQ$};
\end{scope}
}
    
\footnotesize

\begin{circuitikz}
    
    \def\x{1.6}
    \def\y{1.6}
    
    
    
    \coordinate (AN) at (0,0);
    \coordinate (AP) at ($(AN)+(0,\y)$);
    
    \coordinate (FALN) at ($(AN)+(0.5*\x,0)$);
    \coordinate (FALP) at ($(FALN)+(0,\y)$);
    \coordinate (FARN) at ($(FALN)+(\x,0)$);
    \coordinate (FARP) at ($(FARN)+(0,\y)$);
    
    \coordinate (FGLN) at ($(FARN)+(\x,0)$);
    \coordinate (FGLP) at ($(FGLN)+(0,\y)$);
    \coordinate (FGRN) at ($(FGLN)+(\x,0)$);
    \coordinate (FGRP) at ($(FGRN)+(0,\y)$);
    
    \coordinate (GN) at ($(FGRN)+0.5*(\x,0)$);
    \coordinate (GP) at ($(GN)+(0,\y)$);
    
    \draw (FALN)
        to[short] (AN)
        to[voltage source] (AP)
        to[short] (FALP);
    \draw (FALN) to[open,v_=$\VT_{\act,\phsABC}$] (FALP);
    
    \draw (FALN) to[short] (FARN);
    \draw (FALP) to[generic=${R_\act,L_\act}$] (FARP);
    
    \draw (FARN) to[short] (FGLN);
    \draw (FARP)
        to[short,i=$\IT_{\alpha,\phsABC}$] ($0.5*(FARP)+0.5*(FGLP)$)
        to[short] (FGLP); 
    
    \draw ($0.5*(FARN)+0.5*(FGLN)$) to[generic=${G_\fltr,C_\fltr}$,v=$\VT_{\fltr,\phsABC}$,*-*] ($0.5*(FARP)+0.5*(FGLP)$);
    
    \draw (FGLN) to[short] (FGRN);
    \draw (FGLP) to[generic=${R_\grid,L_\grid}$] (FGRP);
    
    \draw (FGRN) to[short,-o] (GN);
    \draw (FGRP) to[short,-o,i=$\IT_{\grid,\phsABC}$] (GP);
    
    \draw (GN) to[open,v^=$\VT_{\grid,\phsABC}$] (GP);
    
    
    
    \coordinate (IA) at ($(AN)-0.1*(0,\y)$);
    \coordinate (OFA) at ($(IA)+(\x,0)$);
    \coordinate (OFI) at ($(OFA)+(\x,0)$);
    \coordinate (OFG) at ($(OFI)+(\x,0)$);
    \coordinate (OG) at ($(OFG)+(\x,0)$);
    
    \transform{TA}{($(IA)-0.8*(0,\y)$)}
    \transform{TFA}{($(TA)+(\x,0)$)}
    \transform{TFI}{($(TFA)+(\x,0)$)}
    \transform{TFG}{($(TFI)+(\x,0)$)}
    \transform{TG}{($(TFG)+(\x,0)$)}
    
    \draw[signal] (TA.north) to node[midway,left]{$\VT_{\act,\phsABC}$} (IA.south);
    \draw[signal] (OFA) to node[midway,left]{$\IT_{\act,\phsABC}$} (TFA.north);
    \draw[signal] (OFI) to node[midway,left]{$\VT_{\fltr,\phsABC}$} (TFI.north);
    \draw[signal] (OFG) to node[midway,left]{$\IT_{\grid,\phsABC}$} (TFG.north);
    \draw[signal] (OG) to node[midway,left]{$\VT_{\grid,\phsABC}$} (TG.north);
    
    
    
    \node[block] (CFA) at ($(TFA)-1.4*(0,\y)$) {\ctrlPI};
    \node[block] (CFI) at ($(CFA)+(\x,0)$) {\ctrlPI};
    \node[block] (CFG) at ($(CFI)+(\x,0)$) {\ctrlPI};
    \node[block] (R) at ($(CFG)+(\x,0)$) {$\div$};
    \node (S) at ($(R)+0.7*(\x,0)$) {$S_\spt$};
    
    \node[dot] (XFI) at ($0.5*(TFI)+0.5*(CFI)$) {};
    \node[dot] (XFG) at ($0.5*(TFG)+0.5*(CFG)$) {};
    \node[dot] (XG) at ($0.5*(TG)+0.5*(R)$) {};
    
    \draw[signal] (TFA.south)
        to node[midway,left]{$\IT_{\act,\cmpDQ}$} ($0.5*(TFA.south)+0.5*(CFA.north)$)
        to (CFA.north);
    \draw[signal] (CFG.west) to node[midway,above]{$\VT^*_{\fltr,\cmpDQ}$} (CFI.east);
    \draw[signal] (CFA.west)
        to ($(CFA)-(\x,0)$)
        to node[near end,left]{$\VT_{\act,\cmpDQ}$} (TA.south);
    
    \draw[-] (TFI.south) to node[midway,left]{$\VT_{\fltr,\cmpDQ}$} (XFI.north);
    \draw[signal] (XFI.south) to (CFI.north);
    \draw[signal] (XFI.west)
        to ($0.5*(TFA.south)+0.5*(CFA.north east)$)
        to ($0.5*(CFA.north)+0.5*(CFA.north east)$);
    \draw[signal] (CFI.west) to node[midway,above]{$\IT^*_{\act,\cmpDQ}$} (CFA.east);

    \draw[-] (TFG.south) to node[midway,left]{$\IT_{\grid,\cmpDQ}$} (XFG.north);
    \draw[signal] (XFG.south) to (CFG.north);
    \draw[signal] (XFG.west) 
        to ($0.5*(TFI.south)+0.5*(CFI.north east)$)
        to ($0.5*(CFI.north)+0.5*(CFI.north east)$);
    
    \draw[-] (TG.south) to node[midway,left]{$\VT_{\grid,\cmpDQ}$} (XG.north);
    \draw[signal] (XG.south) to (R.north);
    \draw[signal] (XG.west)
        to ($0.5*(TFG.south)+0.5*(CFG.north east)$)
        to ($0.5*(CFG.north)+0.5*(CFG.north east)$);
    
    \draw[signal] (R.west) to node[midway,above]{$\IT^{*}_{\grid,\cmpDQ}$} (CFG.east);
    
    \draw[signal] (S.west) to (R.east);
    
\end{circuitikz}

}
	\caption
	{%
		Schematic diagram of a grid-following \CIDER with an LCL filter.
	}
	\label{fig:follow:schem}
\end{figure}


\cref{fig:follow:schem} shows the schematic diagram of a typical grid-following \CIDER.
Its power hardware consists of a \PWM actuator and an LCL filter, and its control software of a three-stage \ctrlPI controller.
The actuator is a three-leg power converter, which is commonly used for grid-following \CIDER[s].



The state of the power hardware is described by the inductor currents $\IT_{\act,\phsABC}\in\RealNum^{3\times1}$ and $\IT_{\grid,\phsABC}\in\RealNum^{3\times1}$ and the capacitor voltage $\VT_{\fltr,\phsABC}\in\RealNum^{3\times1}$.
The input is the actuator voltage $\VT_{\act,\phsABC}\in\RealNum^{3\times1}$ and the disturbance is the grid voltage $\VT_{\grid,\phsABC}\in\RealNum^{3\times1}$.
The output consists of the state and the disturbance.
Formally
\begin{alignat}{2}
	    \XT_\pwr(t)
	&=	\begin{bmatrix}
			\IT_{\act,\phsABC}(t)\\
			\VT_{\fltr,\phsABC}(t)\\
			\IT_{\grid,\phsABC}(t)
		\end{bmatrix}
    &~\in~\RealNum^{9\times1}
	\label{eq:pq:pwr:state}
	\\
	    \UT_\pwr(t)
	&=	\VT_{\act,\phsABC}(t)
    &~\in~\RealNum^{3\times1}
    \\
	    \WT_\pwr(t)
	&=	\VT_{\grid,\phsABC}(t)
    &~\in~\RealNum^{3\times1}
    \\
		\YT_\pwr(t)
	&=	\begin{bmatrix}
	        \XT_\pwr(t)\\
	        \WT_\pwr(t)
	   \end{bmatrix}
    &~\in~\RealNum^{12\times1}
	\label{eq:pq:pwr:output}
\end{alignat}
The matrices of the state-space model are obtained as
\begin{align}
		\AP_\pwr(t)
	&=  \begin{bmatrix}
			-\LP_\act^{-1}\RP_\act	&-\LP_\act^{-1}             &\zero{3}{3}        \\
			\CP_\fltr^{-1}			&-\CP_\fltr^{-1}\GP_\fltr   &-\CP_\fltr^{-1}    \\
			\zero{3}{3}	            &\LP_\grid^{-1}             &-\LP_\grid^{-1}\RP_\grid
		\end{bmatrix}
	\\
		\BP_\pwr(t)
	&=	\begin{bmatrix}
	        \LP_\act^{-1}\\
	        \mathbf{0}_{6\times3}
	    \end{bmatrix}
	\\
	    \EP_{\pwr}(t)
	&=	\begin{bmatrix}
	        \mathbf{0}_{6\times3}\\
	        -\LP_\fltr^{-1}
	    \end{bmatrix}
	\\
        \CP_\pwr(t)
	&=	\begin{bmatrix}
	        \diag(\mathbf{1}_9)\\
	        \mathbf{0}_{3\times9}
	    \end{bmatrix}
	\\
	    \DP_{\pwr}(t)
	&=	\mathbf{0}_{12\times3}
	\\
	    \FP_{\pwr}(t)
	&=	\begin{bmatrix}
	        \mathbf{0}_{9\times3}\\
	        \diag(\mathbf{1}_3)
	    \end{bmatrix}
\end{align}
Their sizes follow straightforward from \eqref{eq:pq:pwr:state}--\eqref{eq:pq:pwr:output}.
Note that these matrices are time-invariant as in the grid-forming case.



Analogously, the state-space variables of the control software are given by
\begin{alignat}{2}
	    \XT_\ctrl(t)
	&=	\int
		\begin{bmatrix}
			\Delta\IT_{\act,\cmpDQ}(t)	\\
			\Delta\VT_{\fltr,\cmpDQ}(t)	\\
			\Delta\IT_{\grid,\cmpDQ}(t)
		\end{bmatrix}
		dt
    &~\in~\RealNum^{6\times1}
	\label{eq:pq:ctrl:state}
	\\
	    \UT_\ctrl(t)
	&=	\begin{bmatrix}
			\IT_{\act,\cmpDQ}(t)	\\
			\VT_{\fltr,\cmpDQ}(t)	\\
			\IT_{\grid,\cmpDQ}(t)   \\
			\VT_{\grid,\cmpDQ}(t)
		\end{bmatrix}
    &~\in~\RealNum^{8\times1}
    \\
	    \WT_\ctrl(t)
	&=	\IT^*_{\grid,\cmpDQ}(t)
    &~\in~\RealNum^{2\times1}
    \\
		\YT_\ctrl(t)
	&=	\VT_{\act,\cmpDQ}(t)
    &~\in~\RealNum^{2\times1}
	\label{eq:pq:ctrl:output}
\end{alignat}
The matrices of the state-space model are obtained as 
\begin{align}
        \AP_\ctrl(t)
    &=  \begin{bmatrix}
                \mathbf{0}_{2\times2}
            &   \frac{\KP_{\ctrlFB,\fltr}}{T_{\ctrlFB,\fltr}}
            &   (\KP_{\ctrlFB,\fltr}+\KP_{\ctrlFF,\fltr})\frac{\KP_{\ctrlFB,\grid}}{T_{\ctrlFB,\grid}}\\
                \mathbf{0}_{2\times2}
            &   \mathbf{0}_{2\times2}
            &   \frac{\KP_{\ctrlFB,\grid}}{T_{\ctrlFB,\grid}}\\
                \mathbf{0}_{2\times2}
            &   \mathbf{0}_{2\times2}
            &   \mathbf{0}_{2\times2}
        \end{bmatrix}
    \\
        \BP_\ctrl(t)
    &=  \begin{bmatrix}
                (\BP_\ctrl)_{11}
            &   -\KP_{\ctrlFB,\fltr}
            &   (\BP_\ctrl)_{13}
            &   (\BP_\ctrl)_{14}\\
               \mathbf{0}_{2\times2}
            &   (\BP_\ctrl)_{22}
            &   -\KP_{\ctrlFB,\grid}
            &   \KP_{\ctrlFT,\grid}\\
                \mathbf{0}_{2\times2}
            &   \mathbf{0}_{2\times2}
            &   (\BP_\ctrl)_{33}
            &   \mathbf{0}_{2\times2}
        \end{bmatrix}\\
		(\BP_\ctrl)_\mathit{ii}
	&=  -\diag(\mathbf{1}_{2}) ~ \forall i\\
	    (\BP_\ctrl)_{13}
	&=  \KP_{\ctrlFT,\fltr} - (\KP_{\ctrlFB,\fltr}+\KP_{\ctrlFF,\fltr})\KP_{\ctrlFB,\grid}\\
	    (\BP_\ctrl)_{14}
	&=  (\KP_{\ctrlFB,\fltr}+\KP_{\ctrlFF,\fltr})\KP_{\ctrlFT,\grid}\\
        \EP_{\ctrl}(t)
	&=	\begin{bmatrix}
	        (\KP_{\ctrlFB,\fltr}+\KP_{\ctrlFF,\fltr})(\KP_{\ctrlFB,\grid}+\KP_{\ctrlFF,\grid})\\
	        \KP_{\ctrlFB,\grid}+\KP_{\ctrlFF,\grid}\\
	        \diag(\mathbf{1}_2)
	    \end{bmatrix}
    \\
        \CP_\ctrl(t)
    &=  \begin{bmatrix}
                \frac{\KP_{\ctrlFB,\act}}{T_{\ctrlFB,\act}}
            &   (\KP_{\ctrlFB,\act}+\KP_{\ctrlFF,\act})\frac{\KP_{\ctrlFB,\fltr}}{T_{\ctrlFB,\fltr}}
            &   (\CP_\ctrl)_3
        \end{bmatrix}\\
        (\CP_\ctrl)_3
    &=  (\KP_{\ctrlFB,\act}+\KP_{\ctrlFF,\act})(\KP_{\ctrlFB,\fltr}+\KP_{\ctrlFF,\fltr})
        \frac{\KP_{\ctrlFB,\grid}}{T_{\ctrlFB,\grid}}
    \\
        \DP_\ctrl(t)
    &=  \begin{bmatrix}
                -\KP_{\ctrlFB,\act}
            &   (\DP_\ctrl)_2
            &   (\DP_\ctrl)_3
            &   (\DP_\ctrl)_4
        \end{bmatrix}\\
        (\DP_\ctrl)_2
    &=  \KP_{\ctrlFT,\act}-(\KP_{\ctrlFB,\act}+\KP_{\ctrlFF,\act})\KP_{\ctrlFB,\fltr}\\
        (\DP_\ctrl)_3
    &=  (\KP_{\ctrlFB,\act}+\KP_{\ctrlFF,\act})\hspace{-1mm}
        \left(\hspace{-1.5mm}
        \begin{aligned}
                &\KP_{\ctrlFT,\fltr}\\
            -   &(\KP_{\ctrlFB,\fltr}+\KP_{\ctrlFF,\fltr})\KP_{\ctrlFB,\grid}
        \end{aligned}\hspace{-0.5mm}
        \right)\hspace{-2mm}\\
        (\DP_\ctrl)_4
    &=  (\KP_{\ctrlFB,\act}+\KP_{\ctrlFF,\act})(\KP_{\ctrlFB,\fltr}+\KP_{\ctrlFF,\fltr})
            \KP_{\ctrlFB,\grid}
    \\
        \FP_\ctrl(t)
	&=  \prod_{\lambda\in\{\act,\fltr,\grid\}}\left\{\KP_{\ctrlFB,\lambda}+\KP_{\ctrlFF,\lambda}\right\}
\end{align}
Their sizes follow straightforward from \eqref{eq:pq:ctrl:state}--\eqref{eq:pq:ctrl:output}.
Note that these matrices are also time-invariant.



In grid-following \CIDER[s], the reference angle $\theta$ needed for the \cmpDQ transform is provided by a synchronization unit, usually a \emph{Phase-Locked Loop} (\PLL).
The reference current $\IT^*_{\grid,\cmpDQ}$ is computed in order to track the power setpoint $S_\spt=P_\spt+jQ_\spt$ at the fundamental frequency.
Without imposing any further conditions on $\theta$, $\IT^*_{\grid,\cmpDQ}$ is given by
\begin{equation}
    \IT^*_{\grid,\cmpDQ}(t)
    =   \frac{1}{v_{\grid,\cmpD}^2(t)+v_{\grid,\cmpQ}^2(t)}\hspace{-0.5mm}
        \begin{bmatrix}
                v_{\grid,\cmpD}(t)
            &   -v_{\grid,\cmpQ}(t)\\
                v_{\grid,\cmpQ}(t)
            &   \phantom{+}v_{\grid,\cmpD}(t)
        \end{bmatrix}\hspace{-1.5mm}
        \begin{bmatrix}
            P_\spt\\
            Q_\spt
        \end{bmatrix}
    \label{eq:ref:following:general}
\end{equation}
In the vast majority of cases, synchronization units in general and \PLL[s] in particular are designed to lock to the fundamental positive-sequence component of the grid voltage (e.g., \cite{Bk:Teodorescu:2011}).
This working principle leads to the following hypothesis:
\begin{Hypothesis} \label{hyp:rsc:follower:synch}
    The synchronization units of the grid-following \CIDER[s] lock to the fundamental positive-sequence component of the grid voltage.
    Therefore, in steady state it holds that
    \begin{equation}
            V_{\grid,\cmpQ,0}
        =   \frac{1}{T}\int v_{\grid,\cmpQ}(t)\,dt
        =   0
    \end{equation}
\end{Hypothesis}
\noindent
For instance, this can be achieved by a closed-loop controller which adjusts $\theta$ in order to regulate $v_{\grid,\cmpQ}$ to 0 on average.

As required by power quality standards (e.g., \cite{Std:BSI-EN-50160:2000}), the grid voltages have to be maintained balanced and sinusoidal within specified limits\footnote{By contrast, the grid currents may be subject to unbalances and harmonics.}.
Under these conditions, it follows that
\begin{Hypothesis}\label{hyp:rsc:follower:harms}
    The time-variant signal content of $v_{\grid,\cmpD}(t)$ and $v_{\grid,\cmpQ}(t)$, as given by $\xi_\cmpD(t)$ and $\xi_\cmpQ(t)$ below, is low:
    \begin{alignat}{2}
                v_{\grid,\cmpD}(t)
        &=      V_{\grid,\cmpD,0}(1+\xi_\cmpD(t)),~
        &       \Abs{\xi_\cmpD(t)}
        &\ll    1\\
                v_{\grid,\cmpQ}(t)
        &=      V_{\grid,\cmpQ,0}(1+\xi_\cmpQ(t)),~
        &       \Abs{\xi_\cmpQ(t)}
        &\ll    1
    \end{alignat}
\end{Hypothesis}
\noindent
As a consequence of \cref{hyp:rsc:follower:synch,hyp:rsc:follower:harms}, $v_{\grid,\cmpQ}(t)$ can be neglected w.r.t. $v_{\grid,\cmpD}(t)$ in \eqref{eq:ref:following:general}:
\begin{equation}
        \IT^*_{\grid,\cmpDQ}(t)
    =   \frac{1}{v_{\grid,\cmpD}(t)}
        \begin{bmatrix}
            P_\spt\\
            Q_\spt
        \end{bmatrix}
    \label{eq:ref:following:simplified}
\end{equation}
In order to calculate the harmonic-domain closed-loop transfer function of the \CIDER, the Fourier coefficients of $\IT^*_{\grid,\cmpDQ}$ are needed.
Unfortunately, the exact expressions which relate the Fourier coefficients of the reciprocal $v^{-1}_{\grid,\cmpD}$ and those of $v_{\grid,\cmpD}$ are complicated \cite{Jrn:Edrei:1953}, and their evaluation is computationally intensive.
However, taking advantage of \cref{hyp:rsc:follower:harms}, the following approximation is made:
\begin{Hypothesis}\label{hyp:rsc:follower:approx}
    For the calculation of the reference current in the grid-following \CIDER[s], the reciprocal of the grid voltage is approximated by a second-order Taylor series
    \begin{equation}
                \frac{1}{v_{\grid,\cmpD}(t)}
        \approx \frac{1}{V_{\grid,\cmpD,0}}\left(1-\xi_\cmpD(t)+\xi_\cmpD^2(t)\right)
    \end{equation}
\end{Hypothesis}
\noindent
Let $\Psi_{h}$ be the Fourier coefficients of the Taylor approximation of the reciprocal $v^{-1}_{\grid,\cmpD}(t)$:
\begin{align}
                \frac{1}{v_{\grid,\cmpD}(t)}
    &\approx    \sum_{h}\Psi_{h}\exp(jh2\pi f_1 t)
    \label{eq:ref:following:invVd}
\end{align}
These Fourier coefficients are obtained by substituting
\begin{equation}
    \xi_{\cmpD}(t) = \sum_{h\neq0}\frac{V_{\grid,\cmpD,h}}{V_{\grid,\cmpD,0}}\Exp{jh2\pi f_1 t}
\end{equation}
into the Taylor series, and expanding the second-order term.
This yields
\begin{align}
        \Psi_{h}
    &=  \left\{
				\begin{array}{cl}
                		\begin{aligned}
                		    \frac{1}{V_{\grid,\cmpD,0}}
					        + \sum_{h\neq0}\frac{\Abs{V_{\grid,\cmpD,h}}^{2}}{V_{\grid,\cmpD,0}^{3}}
					    \end{aligned}
				    &   \text{$h=0$}
				    \\
                		\begin{aligned}
                		    -\frac{V_{\grid,\cmpD,h}}{V_{\grid,\cmpD,0}^2}
					        + \sum_{i\neq0}\frac{V_{\grid,\cmpD,i}V_{\grid,\cmpD,h-i}}{V_{\grid,\cmpD,0}^3}
					    \end{aligned}
				    &   \text{otherwise}
				\end{array}
        \right.
        \label{eq:ref:following:invVd:fourier}
\end{align}
Note well that, since the Taylor series in \cref{hyp:rsc:follower:approx} is of second order, the approximation of the $PQ$ control law \eqref{eq:ref:following:simplified} is a nonlinear function of the Fourier coefficients.


\subsection{Accordance with the Proposed Generic Model}

The previously discussed models are perfectly in accordance with the generic model presented in Part~I of this paper, provided that the listed hypotheses hold.
Therefore, the suitability of the generic model to represent different types of \CIDER[s] can be confirmed at this point.

\section{Validation of the Proposed Method\\ on Individual Resources}
\label{sec:val-rsc}


\subsection{Methodology and Key Performance Indicators}
\label{sec:val-rsc:method}



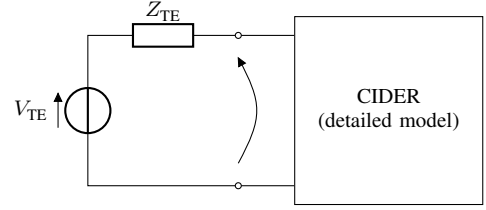
\begin{figure}[t]
	\centering
	{

\footnotesize
\ctikzset{bipoles/length=1.0cm}

\begin{circuitikz}[european]
    
	\def\x{1.0}
	\def\y{1.0}
	
	\coordinate (SN) at (0,0);
	\coordinate (SP) at ($(SN)+(0,2*\y)$);
	\coordinate (ON) at ($(SN)+(2*\x,0)$);
	\coordinate (OP) at ($(SP)+(2*\x,0)$);
	\coordinate (I) at ($0.5*(SN)+0.5*(SP)-(\x,0)$);
	\coordinate (R) at ($(I)+(5*\x,0)$);
	
	
	
	\node[rectangle,draw=black,minimum size=2.5cm] (Res) at (R)
	{%
	    \begin{tabular}{c}
	        \CIDER\\
	        (detailed model)
	    \end{tabular}
	};
	\draw ($(Res.west)+(0,\y)$) to[short] (OP);
	\draw ($(Res.west)+(0,-\y)$) to[short] (ON);
	
	
	\draw (ON)
	    to[short] (SN)
	    to[voltage source = $V_{\TE}$] (SP)
	    to[R = $Z_{\TE}$] (OP);
	\draw (ON) to[open,o-o,v=$ $] (OP);
	
\end{circuitikz}

}
	\caption
	{%
	    Test setup for the validation of the \HPF method on individual \CIDER[s].
	    The resource is represented by a detailed state-space model (see \cref{sec:lib-rsc}), and the power system by a Th{\'e}venin equivalent (see \cref{tab:TE:parameters,tab:TE:harmonics}).
	}
	\label{fig:val-rsc:setup}
\end{figure}

\begin{table}[t]
    \centering
    \caption{Short-Circuit Parameters of the Th{\'e}venin Equivalent}
    \label{tab:TE:parameters}
    {

\renewcommand{\arraystretch}{1.1}

\begin{tabular}{cccl}
    \hline
        Parameter
    &   Resource
    &   System
    &   Description
    \\
    &   Validation
    &   Validation
    \\
    \hline
        $V_{n}$
    &   230\,V-\RMS
    &   230\,V-\RMS
    &   Nominal voltage 
    \\
        $S_{\mathit{sc}}$
    &   267\,kW
    &   3.85\,MW
    &   Short-circuit power
    \\
        $\Abs{Z_{\mathit{sc}}}$
    &   195\,m$\Omega$
    &   13.7\,m$\Omega$
    &   Short-circuit impedance
    \\
        $R_{\mathit{sc}}/X_{\mathit{sc}}$
    &   6.207
    &   0.271
    &   Resistance-to-reactance ratio
    \\
    \hline
\end{tabular}
}
\end{table}

\begin{table}[t]
	\centering
	\caption{Harmonic Voltages of the Th{\'e}venin Equivalent (see \cite{Std:BSI-EN-50160:2000}).}
	\label{tab:TE:harmonics}
	{
\renewcommand{\arraystretch}{1.1}

\begin{tabular}{ccc}
    \hline
        $h$
    &   $|V_{\TE,h}|$
    &   $\angle V_{\TE,h}$
    \\
    \hline
        1
    &   1.0\,p.u.
    &   \phantom{0}0.00\,$\deg$
    \\
        5
    &   6.0\,\%
    &   22.50\,$\deg$
    \\
        7
    &   5.0\,\%
    &   15.00\,$\deg$
    \\ 
        11
    &   3.5\,\%
    &   11.25\,$\deg$
    \\
        13
    &   3.0\,\%
    &   22.50\,$\deg$
    \\
        17
    &   2.0\%
    &   15.00\,$\deg$
    \\
        19
    &   1.5\,\%
    &   11.25\,$\deg$
    \\
        23
    &   1.5\,\%
    &   11.25\,$\deg$
    \\
    \hline
\end{tabular}

}
\end{table}

For the validation of the proposed modelling framework for \CIDER[s], the test setup shown in \cref{fig:val-rsc:setup} is used.
It consists of two parts: a \emph{Th{\'e}venin Equivalent} (\TE) that represents the grid, and a detailed model of the \CIDER under investigation.
The \TE impedance is characterized by typical short-circuit parameters of a power distribution grid, which are given in \cref{tab:TE:parameters}.
The \TE voltage source includes harmonics, whose levels are given in \cref{tab:TE:harmonics}.
These levels are set according to the limits specified in the standard BS-EN-50160:2000 \cite{Std:BSI-EN-50160:2000}.
In line with this standard, harmonics up to order 25 (i.e., 1.25~kHz) are considered in the \HPF study.
Note that, a \TE voltage source with the maximum permissible distortion at each harmonic frequency corresponds to a stressed grid.
This condition is deemed most suitable for the validation of the \HPF method, because it poses a challenge to the modelling framework.
Moreover, it is crucial that the results are reliable when the system is under stress.

The exemplary parameters for the grid-forming and grid-following \CIDER are listed in
\cref{tab:CIDER-forming:parameters,tab:CIDER-following:parameters}, respectively.
The filter parameters were derived following the design rules proposed in \cite{Jrn:Liserre:2005}.
The setpoints are $V_\spt=241.5\,\text{V-\RMS}$ and $f_\spt=50\,\text{Hz}$ for the grid-forming \CIDER, and $P_\spt=50\,\text{kW}$ and $Q_\spt=16.4\,\text{kVAr}$ for the grid-following one.

The \HPF models and method discussed in Part~I of this paper were implemented in Matlab, and compared against \emph{Time-Domain Simulations} (\TDS) with averaged models of the \CIDER[s] in Simulink (recall \cref{hyp:cmp:act}).
The \TDS are stopped once steady-state is reached, and the spectra are calculated using the \emph{Discrete Fourier Transform} (\DFT) on a time window composed by the last 5 periods of the fundamental frequency of the obtained signals.
All analyses are performed in normalized units w.r.t. the base power $P_{b}=10\,\text{kW}$ and the base voltage $V_{b}=230\,\text{V-\RMS}$.

\begin{table}[t]
	\centering
	\caption
	{%
	    Parameters of the Grid-Forming Resource\linebreak
	    (Rated Power $100\,\text{kVA}$)
	}
	\label{tab:CIDER-forming:parameters}
	{

\renewcommand{\arraystretch}{1.1}
\setlength{\tabcolsep}{0.15cm}

\begin{tabular}{lccccc}
	\hline
		Filter stage
	&	$L$/$C$
	&	$R$/$G$
	&	$K_{\ctrlFB}$
	&	$T_{\ctrlFB}$
	&	$K_{\ctrlFT}$
	\\
	\hline
		Inductor ($\alpha$)
	&	0.2\,mH
	&	0.61\,m$\Omega$
	&	15
	&	0.03
	&	1
	\\
	    Capacitor ($\varphi$)
	&	150\,$\mu$F
	&	0\,S
	&	0.05
	&	2.5E-4
	&	0
	\\
	\hline
\end{tabular}

}	
\end{table}

\begin{table}[t]
	\centering
	\caption
	{%
	    Parameters of the Grid-Following Resource\linebreak
	    (Rated Power $60\,\text{kVA}$)
	}
	\label{tab:CIDER-following:parameters}
	{

\renewcommand{\arraystretch}{1.1}
\setlength{\tabcolsep}{0.15cm}

\begin{tabular}{lccccc}
	\hline
		Filter stage
	&	$L$/$C$
	&	$R$/$G$
	&	$K_{\ctrlFB}$
	&	$T_{\ctrlFB}$
	&	$K_{\ctrlFT}$
	\\
	\hline
		Actuator-side inductor ($\alpha$)
	&	325~$\mu$H
	&	1.02~m$\Omega$
	&	10.5
	&	6.6e-4
	&	1
	\\
	    Capacitor ($\varphi$)
	&	90.3~$\mu$F
	&	0~S
	&	1
	&	2.6e-3
	&	0
	\\
		Grid-side inductor ($\gamma$)
	&	325~$\mu$H
	&	1.02~m$\Omega$
	&	0.2
	&	0.1
	&	1
	\\
	\hline
\end{tabular}

}
\end{table}



In order to assess the accuracy and performance of the proposed method, suitable \emph{Key Performance Indicators} (\KPI[s]) have to be defined.
The accuracy is quantified by the errors of the harmonic phasors obtained using the \HPF method w.r.t. the \DFT spectra of the time-domain signals.
Let $\mathbf{X}_{h}$ denote the Fourier coefficient of a polyphase electrical quantity (i.e., voltage or current).
The \KPI[s] are defined as follows:
\begin{align}
                e_{\textup{abs}}(\mathbf{X}_{h})
    &\coloneqq  \max_{p} \Abs{\Abs{X_{h,p,\HPF}}-\Abs{X_{h,p,\TDS}} }\\
                e_{\textup{arg}}(\mathbf{X}_{h})
    &\coloneqq  \max_{p} \Abs{ \angle X_{h,p,\HPF}- \angle X_{h,p,\TDS} }
\end{align}
So, $e_{\textup{abs}}(\mathbf{X}_{h})$ and $e_{\textup{arg}}(\mathbf{X}_{h})$ are the maximum absolute errors in magnitude and phase, respectively, over all phases $p\in\phases$.


\subsection{Results and Discussion}
\label{sec:val-rsc:results}

\begin{figure}[t]
    \centering
    \subfloat[]
    {%
        \centering
        \includegraphics[width=1\linewidth]{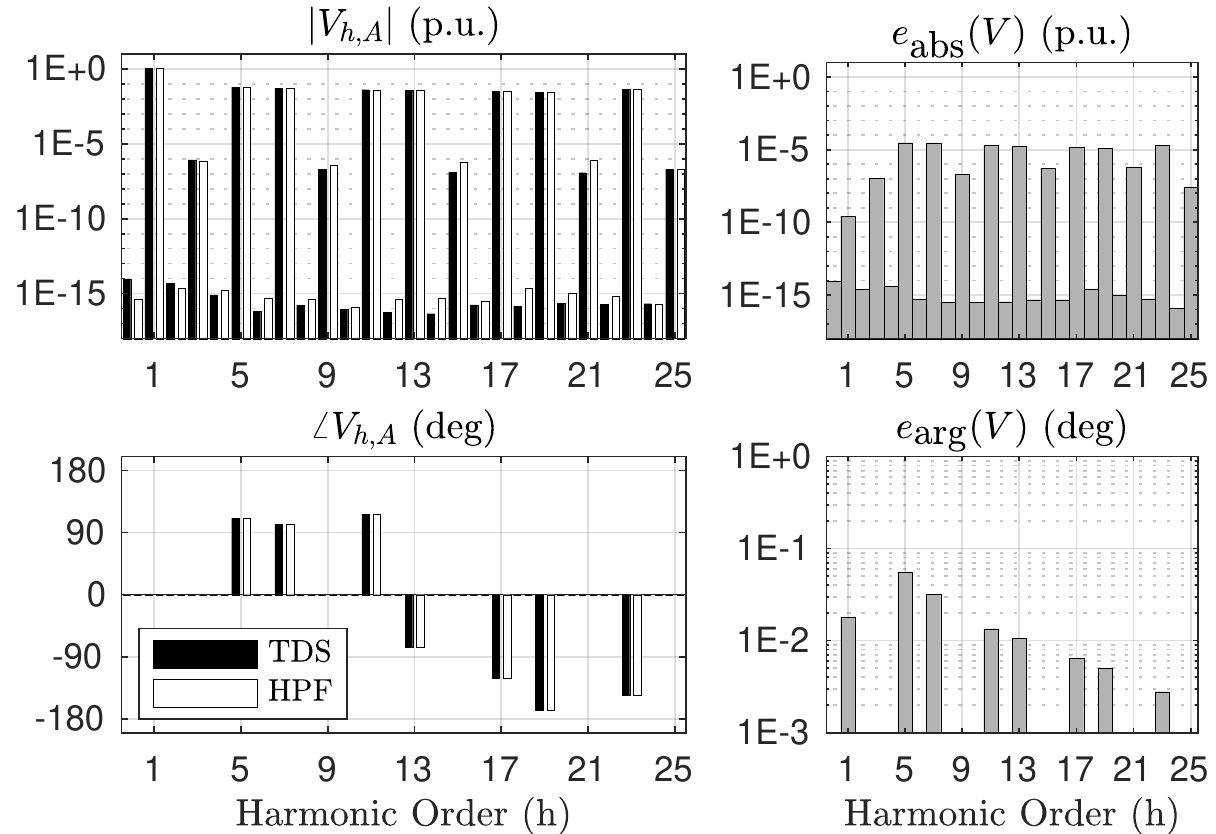}
        \label{fig:resource:results:CIDER-forming}
    }
    
    \subfloat[]
    {%
        \centering
        \includegraphics[width=1\linewidth]{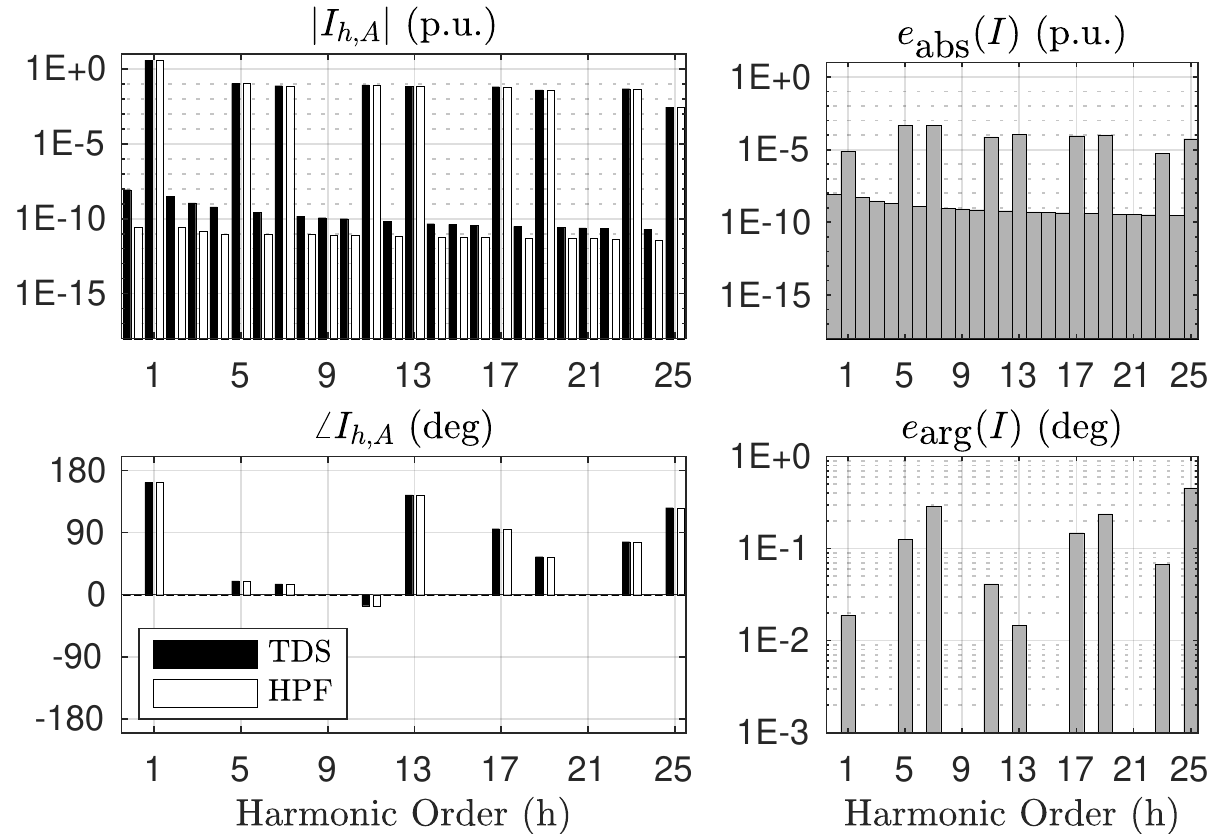}
        \label{fig:resource:results:CIDER-following}
    }
    \caption
    {%
        Results of the validation on individual grid-forming (\ref{fig:resource:results:CIDER-forming}) and grid-following (\ref{fig:resource:results:CIDER-following}) \CIDER[s].
        The plots on the left-hand side show the spectra (i.e., for phase A), and the ones on the right-hand side the error.
    }
    \label{fig:resource:results}
\end{figure}

The result for the grid-forming and grid-following \CIDER are shown in \cref{fig:resource:results:CIDER-forming,fig:resource:results:CIDER-following}, respectively.
For the sake of simplicity, only the controlled quantities are shown: that is, the output voltage of the grid-forming \CIDER, and the output current of the grid-following \CIDER.
The spectra on the left-hand sides of \cref{fig:resource:results:CIDER-forming,fig:resource:results:CIDER-following} show that the Fourier coefficients obtained using \HPF and \TDS are congruent both in magnitude and phase.
This is confirmed by the error plots on the right-hand sides of the figures.
The maximum errors are $e_{\textup{abs}}(\mathbf{V}_{5})=3.01$E-5~p.u. and $e_{\textup{arg}}(\mathbf{V}_{5})=0.06$~deg for the grid-forming resource, and $e_{\textup{abs}}(\mathbf{I}_{7})=4.95$E-4~p.u. and $e_{\textup{arg}}(\mathbf{I}_{25})=0.46$~deg for the grid-following one.
These values are below the accuracy of standard measurement equipment (i.e., they are negligible in practice).


\vfill
\break

\section{Validation of the Proposed Method\\on an Entire System}
\label{sec:val-sys}


\subsection{Methodology and Key Performance Indicators}
\label{sec:val-sys:method}



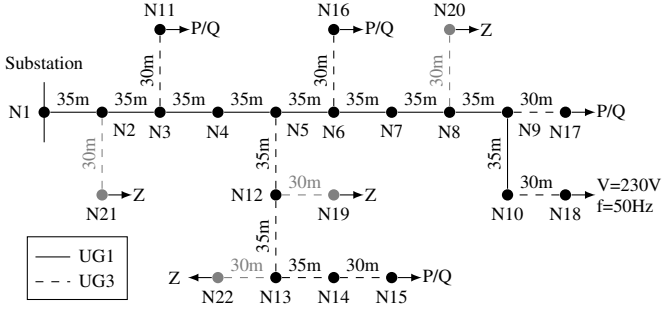
\begin{figure}[t]
	\centering
    {
\tikzstyle{bus}=[circle,fill=black,minimum size=1.5mm,inner sep=0mm]
\tikzstyle{UG1}=[]
\tikzstyle{UG3}=[dashed]
\tikzstyle{load}=[-latex]

\begin{circuitikz}
	\scriptsize
	
	\def\BlockSize{1.0}	
	\def\X{3.8}	
	\def\Y{3.6}	
	\def\dlX{0.77}
	\def\dlY{1.1}
	\def\dload{0.4}
	
	
	\node[bus,label={left:N1}] (N1) at (0,0) {};
	\node[bus,label={below right:N2}] (N2) at (\dlX,0) {};
	\node[bus,gray,label={below:N21}] (N21) at (\dlX,-\dlY) {};
	\node[bus,label={below:N3}] (N3) at (2*\dlX,0) {};
	\node[bus,label={above:N11}] (N11) at (2*\dlX,\dlY) {};
	\node[bus,label={below:N4}] (N4) at (3*\dlX,0) {};
	\node[bus,label={below right:N5}] (N5) at (4*\dlX,0) {};
	\node[bus,label={left:N12}] (N12) at (4*\dlX,-\dlY) {};
	\node[bus,gray,label={below:N19}] (N19) at (5*\dlX,-\dlY) {};
	\node[bus,label={below:N13}] (N13) at (4*\dlX,-2*\dlY) {};
	\node[bus,gray,label={below:N22}] (N22) at (3*\dlX,-2*\dlY) {};
	\node[bus,label={below:N14}] (N14) at (5*\dlX,-2*\dlY) {};
	\node[bus,label={below:N15}] (N15) at (6*\dlX,-2*\dlY) {};
	\node[bus,label={below:N6}] (N6) at (5*\dlX,0) {};
	\node[bus,label={above:N16}] (N16) at (5*\dlX,\dlY) {};
	\node[bus,label={below:N7}] (N7) at (6*\dlX,0) {};
	\node[bus,label={below:N8}] (N8) at (7*\dlX,0) {};
	\node[bus,gray,label={above:N20}] (N20) at (7*\dlX,\dlY) {};
	\node[bus,label={below right:N9}] (N9) at (8*\dlX,0) {};
	\node[bus,label={below:N10}] (N10) at (8*\dlX,-\dlY) {};
	\node[bus,label={below:N17}] (N17) at (9*\dlX,0) {};
	\node[bus,label={below:N18}] (N18) at (9*\dlX,-\dlY) {};

	
	\draw[UG1] (N1) to node[midway,above]{35m} (N2) {};
	\draw[UG1] (N2) to node[midway,above]{35m} (N3) {};
	\draw[UG1] (N3) to node[midway,above]{35m} (N4) {};
	\draw[UG1] (N4) to node[midway,above]{35m} (N5) {};
	\draw[UG1] (N5) to node[midway,above]{35m} (N6) {};
	\draw[UG1] (N6) to node[midway,above]{35m} (N7) {};
	\draw[UG1] (N7) to node[midway,above]{35m} (N8) {};
	\draw[UG1] (N8) to node[midway,above]{35m} (N9) {};
	\draw[UG1] (N10) to node[sloped,anchor=center,above]{35m} (N9) {};
	
	\draw[UG3] (N3) to node[sloped,anchor=center,above]{30m} (N11) {};
	
	\draw[UG3] (N12) to node[sloped,anchor=center,above]{35m} (N5) {};
	\draw[UG3] (N13) to node[sloped,anchor=center,above]{35m} (N12) {};
	\draw[UG3] (N13) to node[midway,above]{35m} (N14) {};
	\draw[UG3] (N14) to node[midway,above]{30m} (N15) {};
	
	\draw[UG3] (N6) to node[sloped,anchor=center,above]{30m} (N16) {};
	\draw[UG3] (N9) to node[sloped,anchor=center,above]{30m} (N17) {};
	\draw[UG3] (N18) to node[sloped,anchor=center,above]{30m} (N10) {};

	\filldraw[gray,UG3] (N12) to node[sloped,anchor=center,above]{30m} (N19) {};
	\draw[gray,UG3] (N8) to node[sloped,anchor=center,above]{30m} (N20) {};
	\draw[gray,UG3] (N21) to node[sloped,anchor=center,above]{30m} (N2) {};
	\draw[gray,UG3] (N22) to node[midway,above]{30m} (N13) {};
	
	\draw[load] (N11) to node[right,align=left]{~P/Q}
	($(N11)+\dload*(1,0)$);
	\draw[load] (N15) to node[right,align=left]{~P/Q}
	($(N15)+\dload*(1,0)$);
	\draw[load] (N16) to node[right,align=left]{~P/Q}
	($(N16)+\dload*(1,0)$);
	\draw[load] (N17) to node[right,align=left]{~P/Q}
	($(N17)+\dload*(1,0)$);
	
	\draw[load] (N19) to node[right,align=left]{~Z}
	($(N19)+\dload*(1,0)$);
	\draw[load] (N20) to node[right,align=left]{~Z}
	($(N20)+\dload*(1,0)$);
	\draw[load] (N21) to node[right,align=left]{~Z}
	($(N21)+\dload*(1,0)$);
	\draw[load] (N22) to node[left,align=right]{Z~}
    ($(N22)-\dload*(1,0)$);

	\draw[load] (N18) to node[right,align=left]{~V=230V\\~f=50Hz}
	($(N18)+\dload*(1,0)$);
	
	\draw[-] ($(N1)+\dload*(0,-1)$) to ($(N1)+\dload*(0,1)$);
	\node[label={above:Substation}] (Substation) at ($(N1)+\dload*(0,1)$) {};

	
    \coordinate (Leg) at (-0.3*\dlX,-1.5*\dlY);
    \matrix [draw,below right] at (Leg) {
        \node [UG1,label=right:~~~~UG1] {}; \\
        \node [UG3,label=right:~~~~UG3] {}; \\
    };
    \draw[UG1] ($(Leg)+(0.15*\dlX,-0.25*\dlY)$) to node[]{} ($(Leg)+(0.75*\dlX,-0.25*\dlY)$);
    \draw[UG3] ($(Leg)+(0.15*\dlX,-0.55*\dlY)$) to node[]{} ($(Leg)+(0.75*\dlX,-0.55*\dlY)$);
    
\end{circuitikz}
}
	\caption
	{%
	    Schematic diagram of the test system, which is based on the \CIGRE low-voltage benchmark microgrid \cite{Rep:2014:CIGRE} (in black) and extended by unbalanced impedance loads (in grey).
	    For the cable parameters see \cref{tab:grid:parameters}.
	    The set of grid-following resources are composed of constant impedance loads (Z) and constant power loads (P/Q), their parameters are given in \cref{tab:resources:references}.
	}
	\label{fig:grid:schematic}
\end{figure}

Lastly, the proposed \HPF method is applied to the test system shown in \cref{fig:grid:schematic}, which is adapted from the \CIGRE low-voltage benchmark microgrid \cite{Rep:2014:CIGRE}.
That is, the \HPF problem is formulated for the complete system model, and solved numerically using the Newton-Raphson method (see Section~V in Part~I).

The test system is characterized as follows.
The substation is located in node N1.
Its short-circuit parameters, which include both the substation transformer and the upstream grid, are listed in \cref{tab:TE:parameters}.
The lines are built from underground cables, whose sequence parameters are given in \cref{tab:grid:parameters}.
Note that, while the proposed \HPF method can treat frequency-dependent cable parameters (see Section~III in Part~I), the parameters of the benchmark microgrid are considered to be frequency-independent.
A preliminary analysis was conducted with \EMTP-RV in order to confirm that this approximation does hold well on the frequency range under consideration (i.e., $\leqslant$1.25\,kHz).
For further details, please see Appendix~\ref{app:cable-parameters}.
Five \CIDER[s] are connected to the ends of the side feeders (i.e., in nodes N11 and N15-18): one grid-forming and four grid-following ones.
Their parameters are the same as for the resource validation, see \cref{tab:CIDER-forming:parameters,tab:CIDER-following:parameters}.
Additionally, unbalanced wye-connected constant-impedance loads, are connected at nodes N19-22.
The unbalance of a load is expressed by phase weights, which indicate the distribution of the load among the phases.
The setpoints and parameters of the grid-following resources are given in \cref{tab:resources:references}, the setpoints of the grid-forming resources in \cref{fig:grid:schematic}.
Notably, the load unbalance is set such that the resulting voltage unbalance does not exceed the limits specified in \cite{Std:BSI-EN-50160:2000} (i.e., $|\mathrm{V}_{1,-}|\leqslant 2\% \cdot |\mathrm{V}_{1,+}|$).

\begin{table}[t]
    \centering
    \caption{Sequence Parameters of the Lines in the Test System.}
    \label{tab:grid:parameters}
	{
\renewcommand{\arraystretch}{1.1}
\setlength{\tabcolsep}{0.15cm}

\begin{tabular}{ccccccc}
    \hline
        ID
    &   $R_{+}/R_{-}$ 
    &   $R_{0}$ 
    &   $L_{+}/L_{-}$ 
    &   $L_{0}$ 
    &   $C_{+}/C_{-}$ 
    &   $C_{0}$ 
    \\
    \hline
        UG1
    &   0.162~$\Omega$
    &   0.529~$\Omega$
    &   0.262~mH
    &   1.185~mH
    &   637~nF
    &   388~nF
    \\
        UG3
    &   0.822~$\Omega$
    &   1.794~$\Omega$
    &   0.270~mH
    &   3.895~mH
    &   637~nF
    &   388~nF
    \\
    \hline
\end{tabular}
}
\end{table}

\begin{table}[t]
    \centering
    \caption{Parameters of the Grid-Following Resources and Loads\newline in the Test System.}
    \label{tab:resources:references}
	{

\renewcommand{\arraystretch}{1.2}
\begin{tabular}{ccccc}
	\hline
		Node
	&	S
	&	pf
	&	Type
	&   Phase weights
	\\
	\hline
		N11
	&	15.0~kW
	&	0.95
	&   P/Q
	&   [0.33 0.33 0.33]
	\\
		N15
	&	52.0~kW
	&	0.95
	&   P/Q
	&   [0.33 0.33 0.33]
	\\
		N16
	&	55.0~kW
	&	0.95
	&   P/Q
	&   [0.33 0.33 0.33]
	\\
		N17
	&	35.0~kW   
	&	0.95
	&   P/Q
	&   [0.33 0.33 0.33]
	\\
		N19
	&	51.2~kW   
	&	0.95
	&   Z
	&   [0.31 0.50 0.19]
	\\
		N20
	&	51.7~kW   
	&	0.95
	&   Z
	&   [0.45 0.23 0.32]
	\\
		N21
	&	61.5~kW   
	&	0.95
	&   Z
	&   [0.24 0.39 0.37]
	\\
		N22
	&	61.9~kW   
	&	0.95
	&   Z
	&   [0.31 0.56 0.13]
	\\
	\hline
\end{tabular}
}
\end{table}



As stated in Part~I of this paper, the \HPF problem is a system of nonlinear equations, which are solved numerically by means of the Newton-Raphson method.
Due to its nonlinearity, the \HPF problem may have multiple solutions, one or several of which may not even be physically meaningful.
As known from numerical analysis, the convergence of an iterative numerical solver to a particular solution can be affected by the choice of the initial point.
Therefore, it is crucial to verify whether multiplicity of solutions occurs, and whether the convergence is robust w.r.t. the initial point.
In order to assess the convergence behaviour of the proposed \HPF method, the initial spectra of voltages and currents are varied.
More precisely, the initial spectra are obtained as a superposition of random positive, negative, and homopolar sequences at each frequency (i.e., fundamental and harmonics), whose magnitudes and phases are uniformly distributed in the intervals $[0,10]\,\text{p.u.}$ and $[0,2\pi]\,\text{rad}$, respectively.



The accuracy of the \HPF method is assessed by means of the same \KPI[s] used for the individual resources: the magnitude and phase errors of the \HPF results w.r.t. \DFT spectra of time-domain waveforms.
A detailed performance analysis of the \HPF study is conducted.
To this end, the method's performance is quantified by the mean and standard deviation of the execution time of the \HPF study through $N=50$ simulations and compared to the execution time of the \TDS (incl. the Fourier analysis).
Moreover, a scalability analysis w.r.t. the numbers of \CIDER[s] and w.r.t. the considered harmonic order in the \HPF is performed.
In the first case, the grid-following \CIDER[s] at nodes N15-17 are consecutively replaced by wye-connected, balanced impedance loads, whose nominal power is equal to the setpoint of the associated \CIDER.
In the second case, the timing analysis for the \HPF is repeated while increasing the considered maximum harmonic order $h_{max}$ consecutively from 11 to 25.
It is important to note that the \TDS takes some time to reach the steady state.
The settling time of this transient analysis strongly depends on the initialization of the simulation.
In order to ensure a fair comparison between the \HPF and the \TDS, the execution time of the latter is measured only for 5 periods in steady state (i.e., the window length required for the \DFT) plus the Fourier analysis (i.e., the \DFT).
Note well that this corresponds to the minimum amount of simulation time which would have to be done even if the initialization of the \TDS were perfect.


\subsection{Results and Discussion}
\label{sec:val-sys:results}



\cref{tab:system:sequences} gives the voltage and current sequence components of the nodes where resources are connected.
Indeed the passive impedance loads at N19-22 introduce significant unbalances in the nodal phase-to-ground voltages and injected currents.



The convergence of the method appears to be robust w.r.t. a random choice of the initial point (i.e., sequence components whose magnitudes and phases are uniformly distributed in  the intervals $[0,10]\,\text{p.u.}$ and $[0,2\pi]\,\text{rad}$, respectively).
In fact, the method always converged to the same solution irrespective of its initial value.
That is, neither divergence of the algorithm nor multiplicity of solutions have been observed.
Naturally, this empirical evidence does not provide a general guarantee.
Nevertheless, the fact that the convergence is not affected even by substantial variations of the initial point speaks for the robustness of the proposed method.



\begin{figure}[t]
    \centering
    \includegraphics[width=1\linewidth]{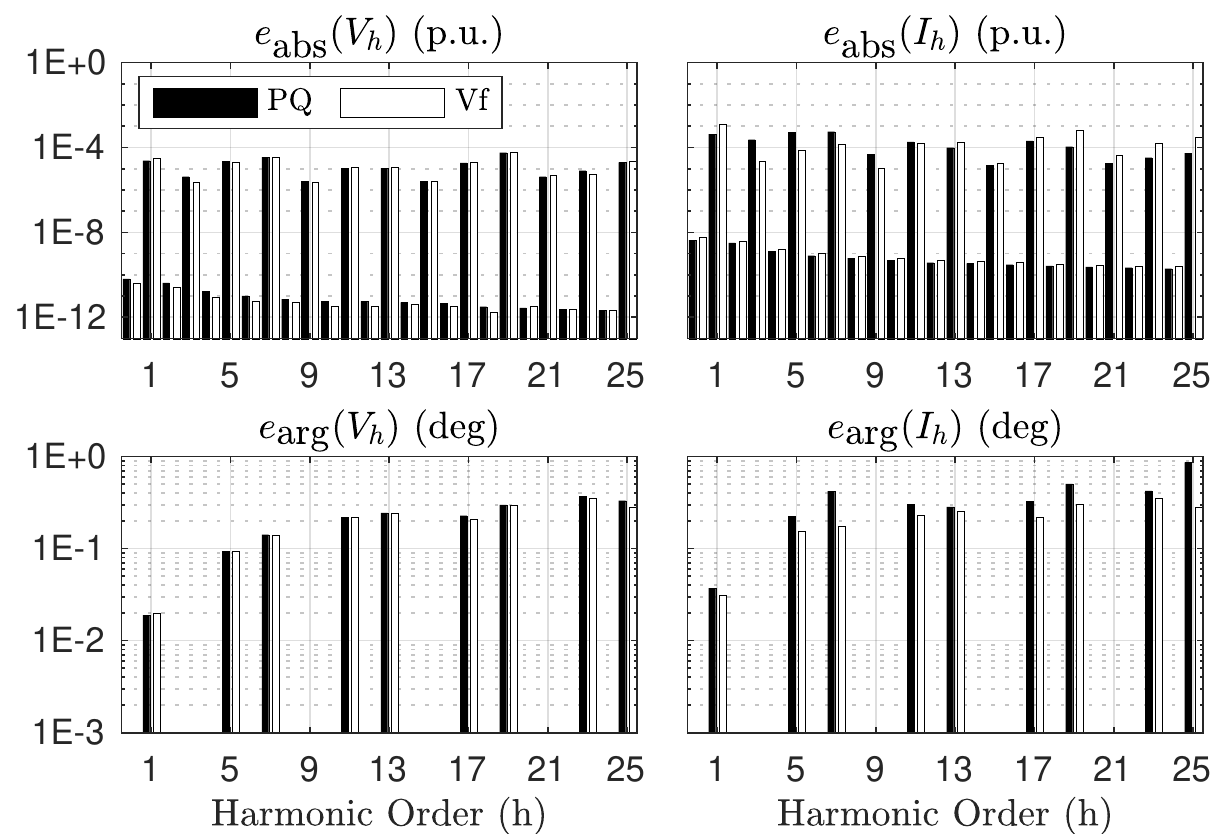}
    \caption
    {%
        Results of the validation on the benchmark system.
        The plots show the maximum absolute errors over all nodes and phases, for voltages (left column) and currents (right column), in magnitude (top row) and phase (bottom row).
    }
    \label{fig:system:error:VI}
\end{figure}

\cref{fig:system:error:VI} shows the maximum absolute errors over all nodes and phases, separately for grid-forming and grid-following \CIDER[s].
The highest errors w.r.t. voltage magnitude and phase are $e_{\textup{abs}}(\mathbf{V}_{19})=6.33$E-5~p.u. and $e_{\textup{arg}}(\mathbf{V}_{23})=0.37$~deg, respectively.
The highest errors w.r.t. current magnitude and phase are $e_{\textup{abs}}(\mathbf{I}_{1})=1.33$E-3~p.u. and $e_{\textup{arg}}(\mathbf{I}_{25})=0.87$~deg, respectively.
Observe that the magnitude errors of the current harmonics are higher than those of the voltage harmonics, which is likely due to the Taylor approximation in the reference calculation of the grid-following \CIDER[s] (i.e., \cref{hyp:rsc:follower:approx}).
Moreover, note that the phase error becomes slightly larger as the harmonic order increases.
Nevertheless, the error levels are generally very low.
Indeed, as it was the case in the resource validation, the magnitude and phase errors are lower than the accuracy of standard measurement equipment.



\begin{figure}[tb]
    \centering
    \includegraphics[width=1\linewidth]{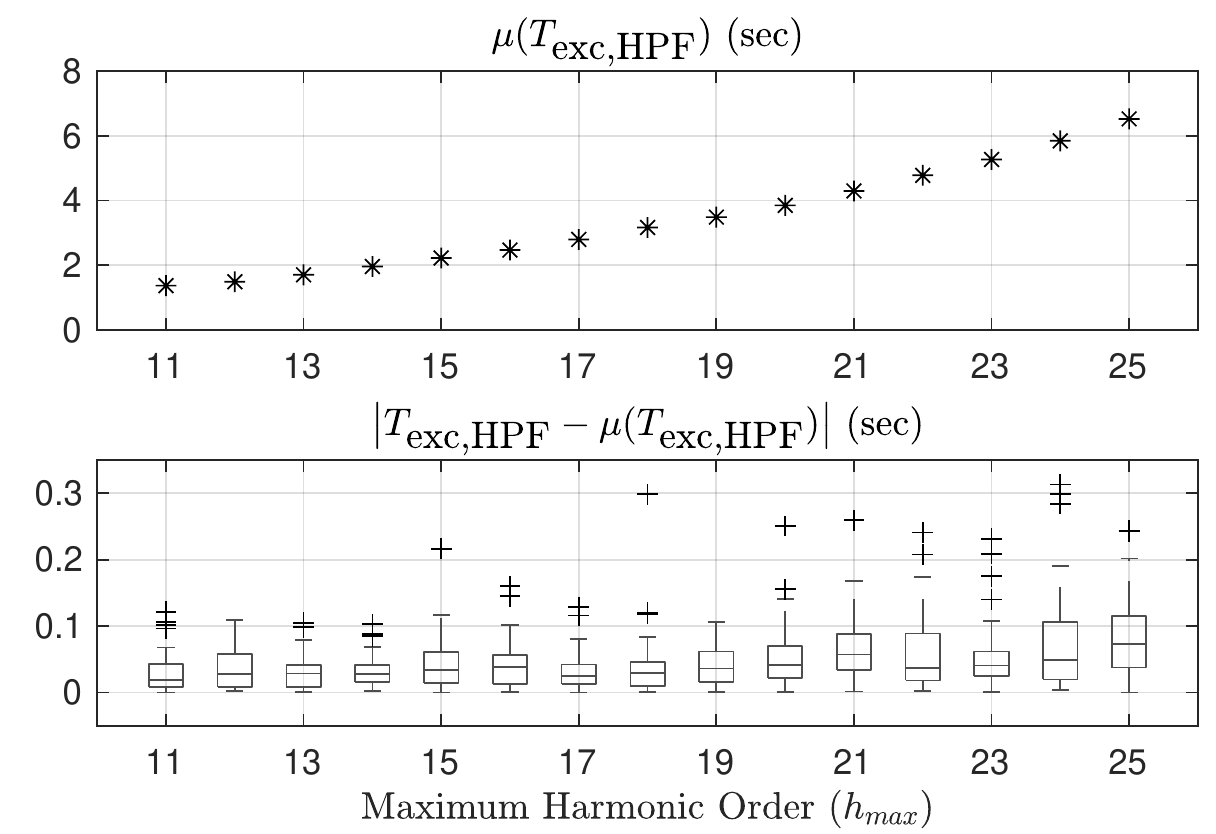}
    \caption
    {%
        Mean and distribution of the timing performance of \HPF for maximum numbers of \CIDER[s] and with varying $h_{max}$ for $N=50$ simulations.
        The box-and-whisker plot visualizes 25 and 75 percentile of the sample, the whisker length is 1.5-times the interquartile range.
    }
    \label{fig:system:timing_hmax}
\end{figure}

All simulations are run on the same laptop computer, namely a MacBook Pro 2019 with a 2.4 GHz Intel Core i9 CPU and 32 GB 2400 MHz DDR3 RAM.
As shown in \cref{tab:system:timing}, the mean of the execution time of the \HPF method lies between 1.75-6.52~sec with standard deviations from 0.03-0.1~sec depending on the number of \CIDER[s] that are connected.
By comparison, the execution time of the \TDS is around 28.33-33.39~sec, out of which ca. 0.6~sec are needed for the Fourier analysis (i.e., the \DFT).
Clearly, the \HPF method is faster than the \TDS, while yielding accurate results.
The computational complexity of the \HPF method in function of the maximum harmonic order is illustrated in the upper subplot of \cref{fig:system:timing_hmax}.
Note that the execution time increases almost linearly, but a non-dominant higher-order component is clearly visible (i.e., as expected based on the involved matrix operations).
The non-deterministic component of $T_{exc,\HPF}$ (i.e., the variation around the mean value) is illustrated in the lower subplot of \cref{fig:system:timing_hmax}.
Observe that any deviation is small compared to the actual value of $T_{exc,\HPF}$.

\begin{table}[t]
    \centering
    \caption{Ratio of Sequence Voltages and Currents\linebreak at the Nodes with Resources}
    \label{tab:system:sequences}
	{

\renewcommand{\arraystretch}{1.1}
\setlength{\tabcolsep}{0.15cm}

\begin{tabular}{lcccc}
	\hline
		Node
	&   $\frac{\Abs{\mathrm{V}_{1,-}}}{\Abs{\mathrm{V}_{1,+}}}$
	&   $\frac{\Abs{\mathrm{V}_{1,0}}}{\Abs{\mathrm{V}_{1,+}}}$
	&   $\frac{\Abs{\mathrm{I}_{1,-}}}{\Abs{\mathrm{I}_{1,+}}}$
	&   $\frac{\Abs{\mathrm{I}_{1,0}}}{\Abs{\mathrm{I}_{1,+}}}$
	\\
	\hline
	    N1
	&   0.26~\%
	&   0.19~\%
	&   12.78~\%
	&   \phantom{0}9.04~\%
	\\
	    N11
	&   0.46~\%
	&   0.77~\%
	&   \phantom{0}0.24~\%
	&   \phantom{0}0.00~\%
	\\
	    N15
	&   1.63~\%
	&   3.76~\%
	&   \phantom{0}0.32~\%
	&   \phantom{0}0.00~\%
	\\
	    N16
	&   0.53~\%
	&   0.80~\%
	&   \phantom{0}0.11~\%
	&   \phantom{0}0.00~\%
	\\
	    N17
	&   0.48~\%
	&   0.51~\%
	&   \phantom{0}0.13~\%
	&   \phantom{0}0.00~\%
	\\
	    N18
	&   0.47~\%
	&   0.23~\%
	&   \phantom{0}0.14~\%
	&   \phantom{0}3.92~\%
	\\
	    N19
	&   1.38~\%
	&   3.17~\%
	&   26.19~\%
	&   24.45~\%
	\\
	    N20
	&   0.40~\%
	&   0.77~\%
	&   19.67~\%
	&   18.90~\%
	\\
	    N21
	&   0.48~\%
	&   0.61~\%
	&   13.18~\%
	&   13.38~\%
	\\
	    N22
	&   1.95~\%
	&   4.57~\%
	&   37.09~\%
	&   34.36~\%
	\\
	\hline
\end{tabular}

}
\end{table}

\begin{table}[t]
    \centering
    \caption{Timing Performance (for $N=50$ Simulations of the \HPF)}
    \label{tab:system:timing}
	{

\renewcommand{\arraystretch}{1.1}
\setlength{\tabcolsep}{0.15cm}

\begin{tabular}{lcccc}
	\hline
		No. of Following \CIDER[s]
	&	1
	&	2
	&	3
	&	4
	\\
	\hline
		$T_{exc,\TDS}$
	&	28.33~sec
	&	30.36~sec
	&	31.28~sec
	&	33.39~sec
	\\
		$\mu(T_{exc,\HPF})$
	&	\phantom{0}1.75~sec
	&	\phantom{0}3.97~sec
	&	\phantom{0}5.30~sec
	&	\phantom{0}6.52~sec
	\\
		$\sigma(T_{exc,\HPF})$
	&	\phantom{0}0.03~sec
	&	\phantom{0}0.07~sec
	&	\phantom{0}0.08~sec
	&	\phantom{0}0.10~sec
	\\
	\hline
\end{tabular}

}
\end{table}

\section{Conclusions}
\label{sec:concl}

In Part~II of this paper, the \HPF method proposed in Part~I was validated.
It was confirmed that common types of \CIDER[s] (i.e., power converters with LC and LCL filters) can well be represented using the proposed modelling framework.
Moreover, the \HPF method was implemented in Matlab and validated against time-domain simulations with Simulink.
For this validation, both individual resources as well as an entire system (i.e., the \CIGRE low-voltage benchmark microgrid) were investigated.
The largest observed errors are 1.33E-3~p.u. w.r.t. current magnitude, 6.33E-5~p.u. w.r.t. voltage magnitude, and 0.87~deg w.r.t. phase.
If run on a standard laptop computer, the execution of the \HPF method for the benchmark system is up to five times faster as compared to the \TDS (incl. the Fourier analysis).
These results confirm that the proposed approach is indeed accurate and computationally efficient.



\appendices

\section{Frequency-Dependent Cable Parameters}
\label{app:cable-parameters}


\subsection{Impact on the Branch Admittance of the Line Model}

\begin{figure}[ht]
    \centering
    \includegraphics[width=0.8\linewidth]{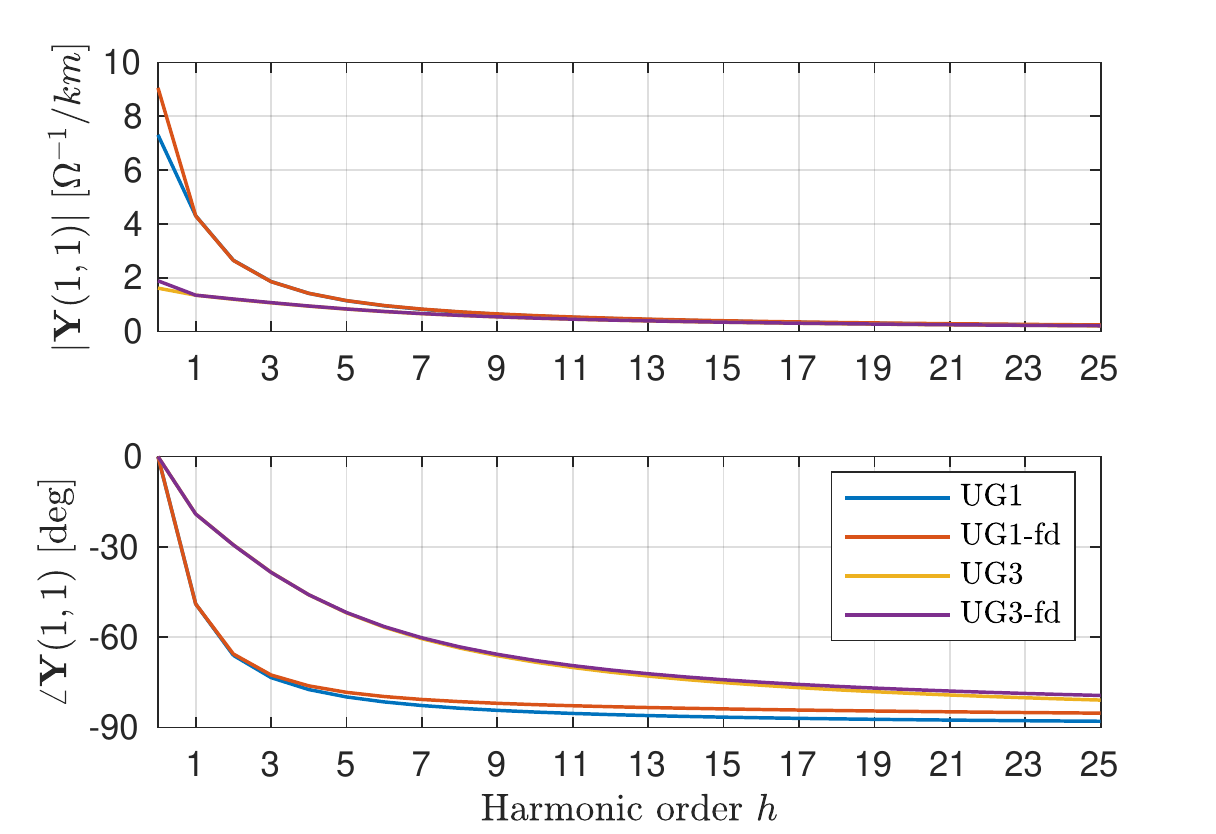}
    \caption{%
        Comparison of line branch admittances of the $\pi$-section equivalents with/without frequency-dependence of the cable parameters.
        For illustration, the element $(1,1)$ of the compound admittance matrices of the cable types (i.e., UG1 and UG3) is shown.
        The curves labelled with the suffix ``fd'' correspond to the cable models with frequency-dependent parameters.
    }
    \label{fig:line-model}
\end{figure}

The \CIGRE report \cite{Rep:2014:CIGRE}, in which the benchmark microgrid is specified, does not provide any information on the frequency dependency of the cable parameters.
Therefore, the cable parameters were calculated using EMTP-RV based on the available data on cable material and geometry.
The behaviour of the branch admittances as function of frequency is shown in \Cref{fig:line-model}.
For illustration, the element $(1,1)$ of the compound admittance matrices are shown.
As one can see, whether or not the frequency dependency of the parameters is considered has virtually no impact on the magnitude of the line admittance.
In the phase, there is a slight difference between the two models at higher frequencies.
        

\subsection{Impact on the Results of the Harmonic Power-Flow Study}

\begin{figure}[ht]
    \centering
    
    \subfloat[]
    {%
        \centering
        \includegraphics[width=\linewidth]{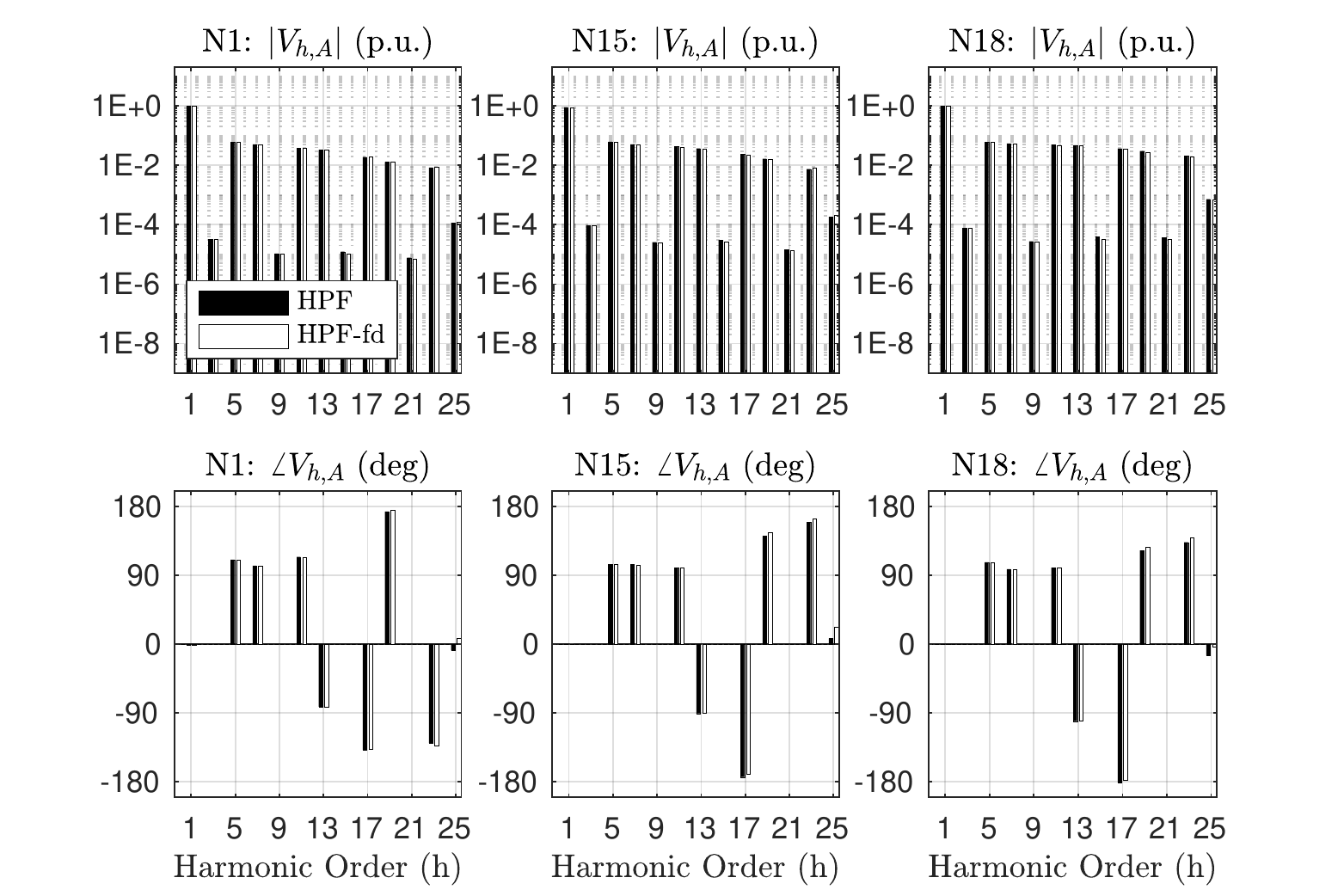}
        \label{fig:results:voltage}
    }
    
    \subfloat[]
    {%
        \centering
        \includegraphics[width=\linewidth]{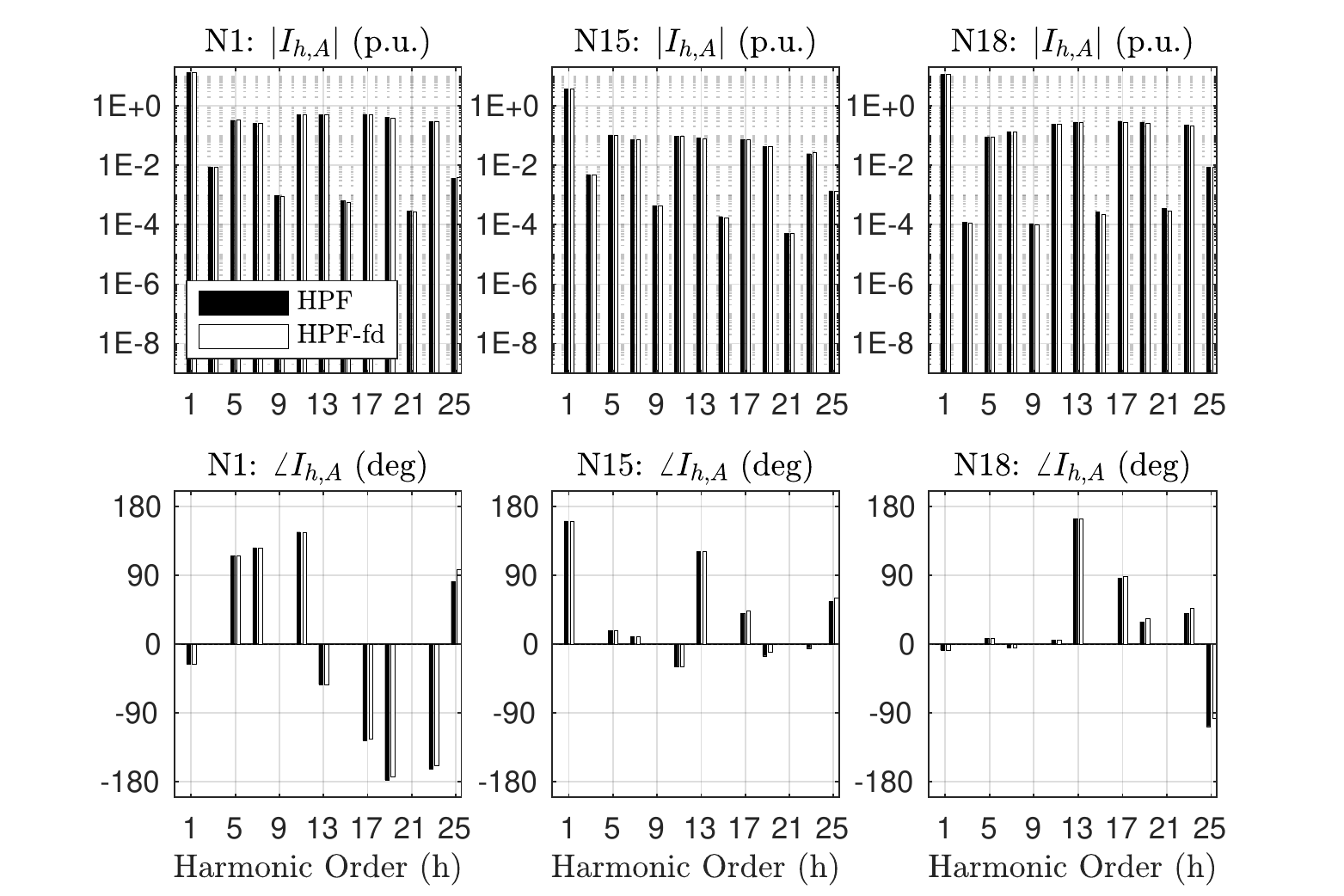}
        \label{fig:results:current}
    }
    
    \caption{%
        Impact of the frequency-dependent parameters on the results of the \HPF study.
        The results are compared at three nodes throughout the benchmark system.
        The currents for Phase A are given in (\ref{fig:results:current}) and the voltages in (\ref{fig:results:voltage}).
        The results labelled with the suffix ``fd'' correspond to the line models with frequency-dependent parameters.
    }
    \label{fig:results}
\end{figure}

In order to assess the impact on the results of the \HPF study, analyses were conducted on the benchmark system using either the line model with frequency-invariant or -dependent parameters.
The obtained results are shown in \Cref{fig:results}.
For illustration, the comparison is done at three nodes throughout the benchmark system.
The results are compared at three nodes throughout the system.
As one can see, the spectra obtained using the different line models are congruent.
This is in line with the previously discussed analyses made in EMTP-RV.



\bibliographystyle{IEEEtran}
\bibliography{Bibliography}

\begin{thebibliography}{10}
\providecommand{\url}[1]{#1}
\csname url@samestyle\endcsname
\providecommand{\newblock}{\relax}
\providecommand{\bibinfo}[2]{#2}
\providecommand{\BIBentrySTDinterwordspacing}{\spaceskip=0pt\relax}
\providecommand{\BIBentryALTinterwordstretchfactor}{4}
\providecommand{\BIBentryALTinterwordspacing}{\spaceskip=\fontdimen2\font plus
\BIBentryALTinterwordstretchfactor\fontdimen3\font minus
  \fontdimen4\font\relax}
\providecommand{\BIBforeignlanguage}[2]{{%
\expandafter\ifx\csname l@#1\endcsname\relax
\typeout{** WARNING: IEEEtran.bst: No hyphenation pattern has been}%
\typeout{** loaded for the language `#1'. Using the pattern for}%
\typeout{** the default language instead.}%
\else
\language=\csname l@#1\endcsname
\fi
#2}}
\providecommand{\BIBdecl}{\relax}
\BIBdecl

\bibitem{Std:BSI-EN-50160:2000}
``Voltage characteristics of electricity supplied by public distribution
  networks,'' British Standards Institution, London, UK, Std. BS-EN-50160:2000,
  2000.

\bibitem{Jrn:PSE:PEC:2004:Blaabjerg}
F.~Blaabjerg, Z.~Chen, and S.~B. Kjaer, ``Power electronics as efficient
  interface in dispersed power generation systems,'' \emph{IEEE Trans. Power
  Electron.}, vol.~19, no.~5, pp. 1184--1194, 2004.

\bibitem{Jrn:Blaabjerg:2006}
F.~Blaabjerg \emph{et~al.}, ``Overview of control and grid synchronization for
  distributed power generation systems,'' \emph{IEEE Trans. Ind. Electron.},
  vol.~53, no.~5, pp. 1398--1409, 2006.

\bibitem{Jrn:Wang:2015}
J.~Wang \emph{et~al.}, ``Design of a generalized control algorithm for parallel
  inverters for smooth microgrid transition operation,'' \emph{IEEE Trans. Ind.
  Electron.}, vol.~62, no.~8, pp. 4900--4914, 2015.

\bibitem{Jrn:Rocabert:2012}
J.~Rocabert \emph{et~al.}, ``Control of power converters in {AC} microgrids,''
  \emph{IEEE Trans. Power Electron.}, vol.~27, no.~11, pp. 4734--4749, 11 2012.

\bibitem{Jrn:Park:1929}
R.~H. Park, ``Two-reaction theory of synchronous machines,'' \emph{Trans.
  AIEE}, vol.~48, no.~3, pp. 716--727, 7 1929.

\bibitem{Bk:Teodorescu:2011}
R.~Teodorescu \emph{et~al.}, \emph{Grid Converters for Photovoltaic and Wind
  Power Systems}.\hskip 1em plus 0.5em minus 0.4em\relax John Wiley \& Sons,
  2011, vol.~29.

\bibitem{Jrn:Edrei:1953}
A.~Edrei and G.~Szeg{\"o}, ``A note on the reciprocal of a {Fourier} series,''
  \emph{Proc. AMS}, vol.~4, no.~2, pp. 323--329, 1953.

\bibitem{Jrn:Liserre:2005}
M.~Liserre, F.~Blaabjerg, and S.~Hansen, ``Design and control of an
  {LCL}-filter-based three-phase active rectifier,'' \emph{IEEE Trans. Ind.
  Appl.}, vol.~41, no.~5, pp. 1281--1291, 2005.

\bibitem{Rep:2014:CIGRE}
K.~Strunz \emph{et~al.}, ``Benchmark systems for network integration of
  renewable and distributed energy resources,'' CIGR{\'E}, Paris, IDF, FR,
  Tech. Rep. 575, 2014.

\end{thebibliography}


\begin{IEEEbiography}[{\includegraphics[width=1in,height=1.25in,clip,keepaspectratio]{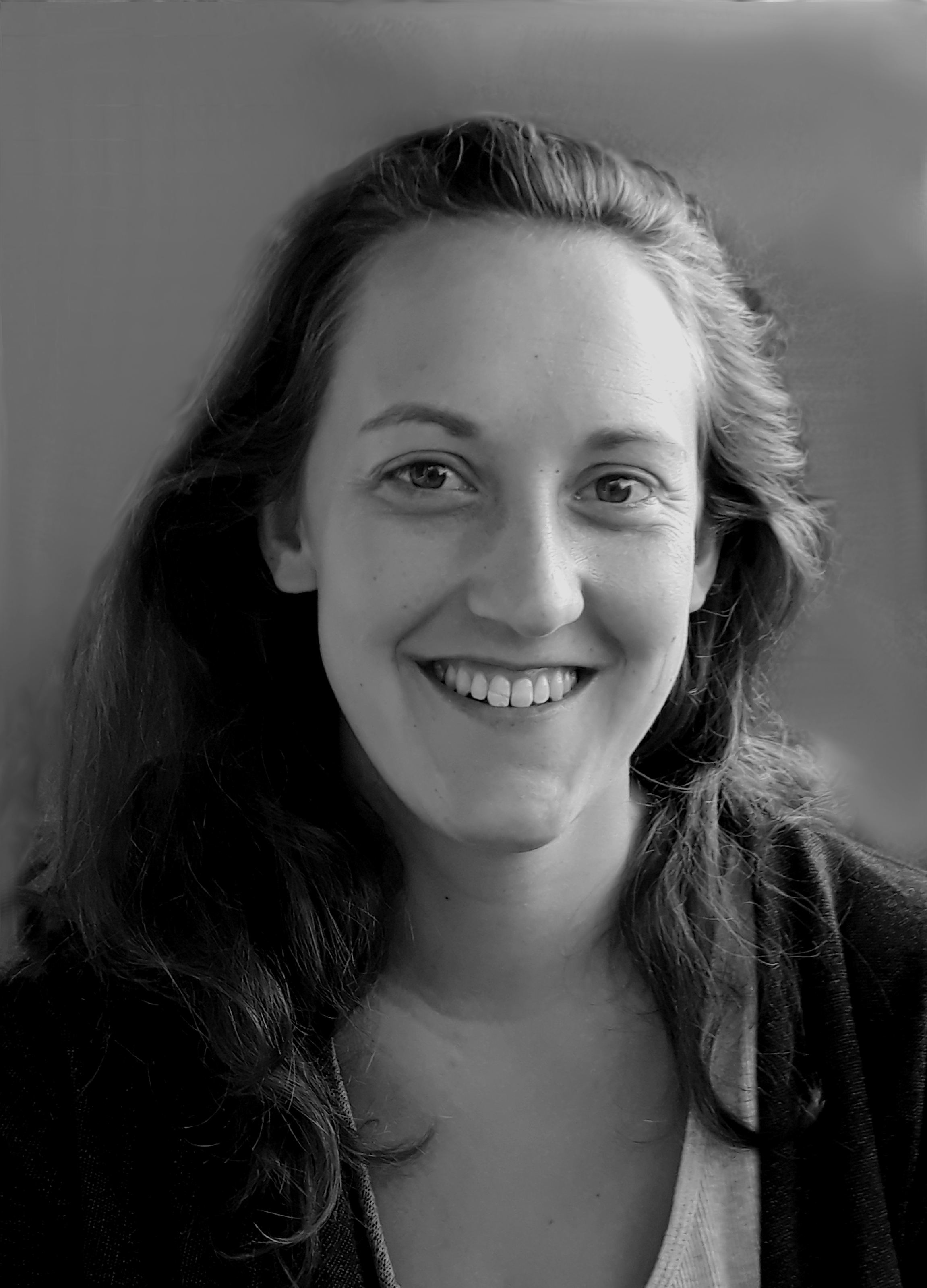}}]{Johanna Kristin Maria Becker}
	(S’19) received the B.Sc. degree in Microsystems Engineering from Freiburg University, Germany in 2015 and the M.Sc. degree in Electrical Engineering from the Swiss Federal Institute of Technology of Lausanne (EPFL), Lausanne, Switzerland in 2019.
    She is currently pursuing a Ph.D. degree at the Distributed Electrical System Laboratory, EPFL, with a focus on robust control and stability assessment of active distribution systems in presence of harmonics.
\end{IEEEbiography}

\begin{IEEEbiography}[{\includegraphics[width=1in,height=1.25in,clip,keepaspectratio]{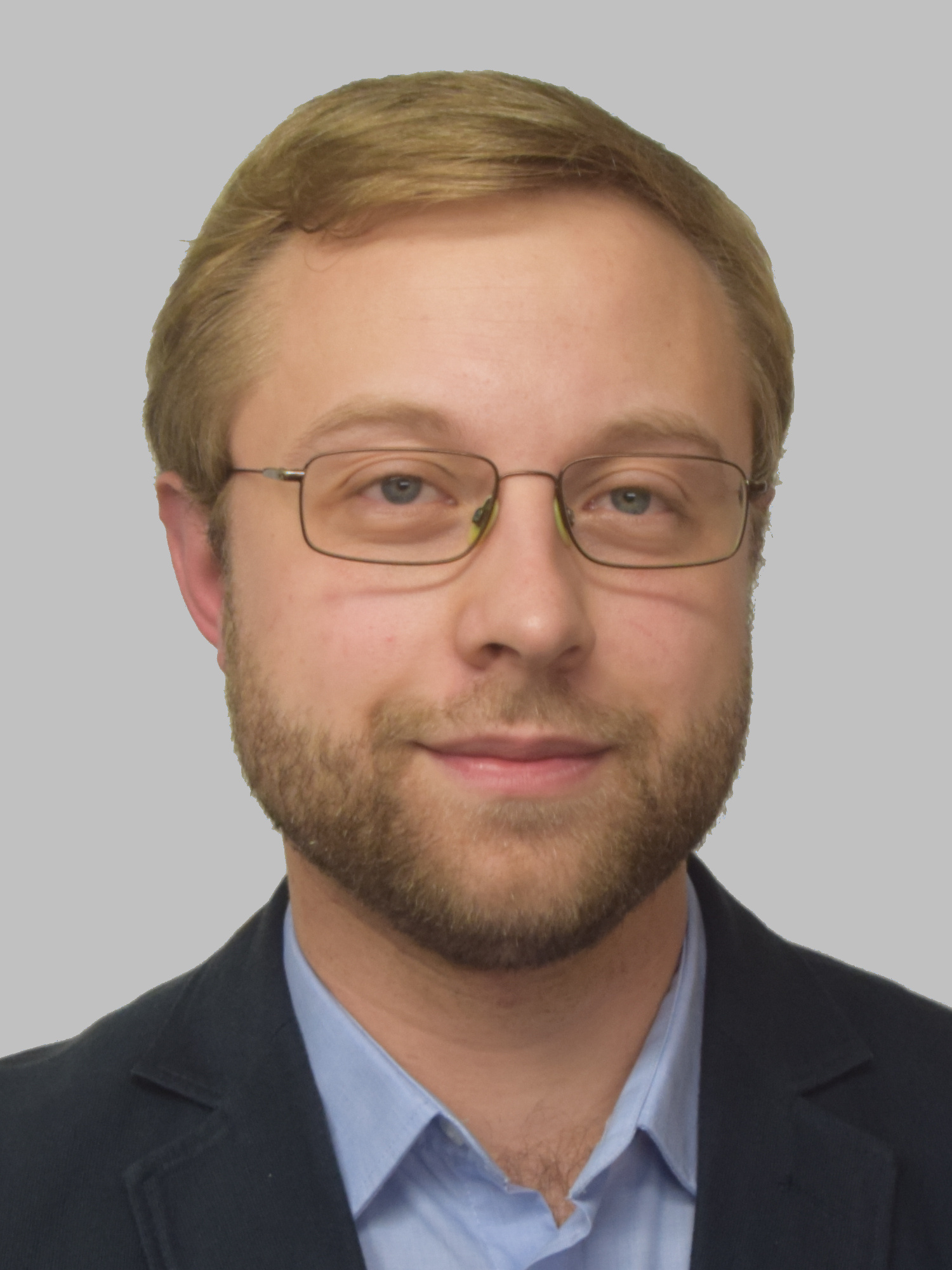}}]{Andreas Martin Kettner}
	(S’15-M’19) received the M.Sc. degree in electrical engineering and information technology from the Swiss Federal Institute of Technology of Zürich (ETHZ), Zürich, Switzerland in 2014, and the Ph.D. degree in power systems engineering from the Swiss Federal Institute of Technology of Lausanne (EPFL), Lausanne, Switzerland in 2019.
	He was a postdoctoral researcher at the Distributed Electrical Systems Laboratory (DESL) of EPFL from 2019 to 2020.
	Since then, he has been working as a project engineer for PSI NEPLAN AG in K{\"u}snacht, Switzerland and continues to collaborate with DESL.
\end{IEEEbiography}

\begin{IEEEbiography}[{\includegraphics[width=1in,height=1.25in,clip,keepaspectratio]{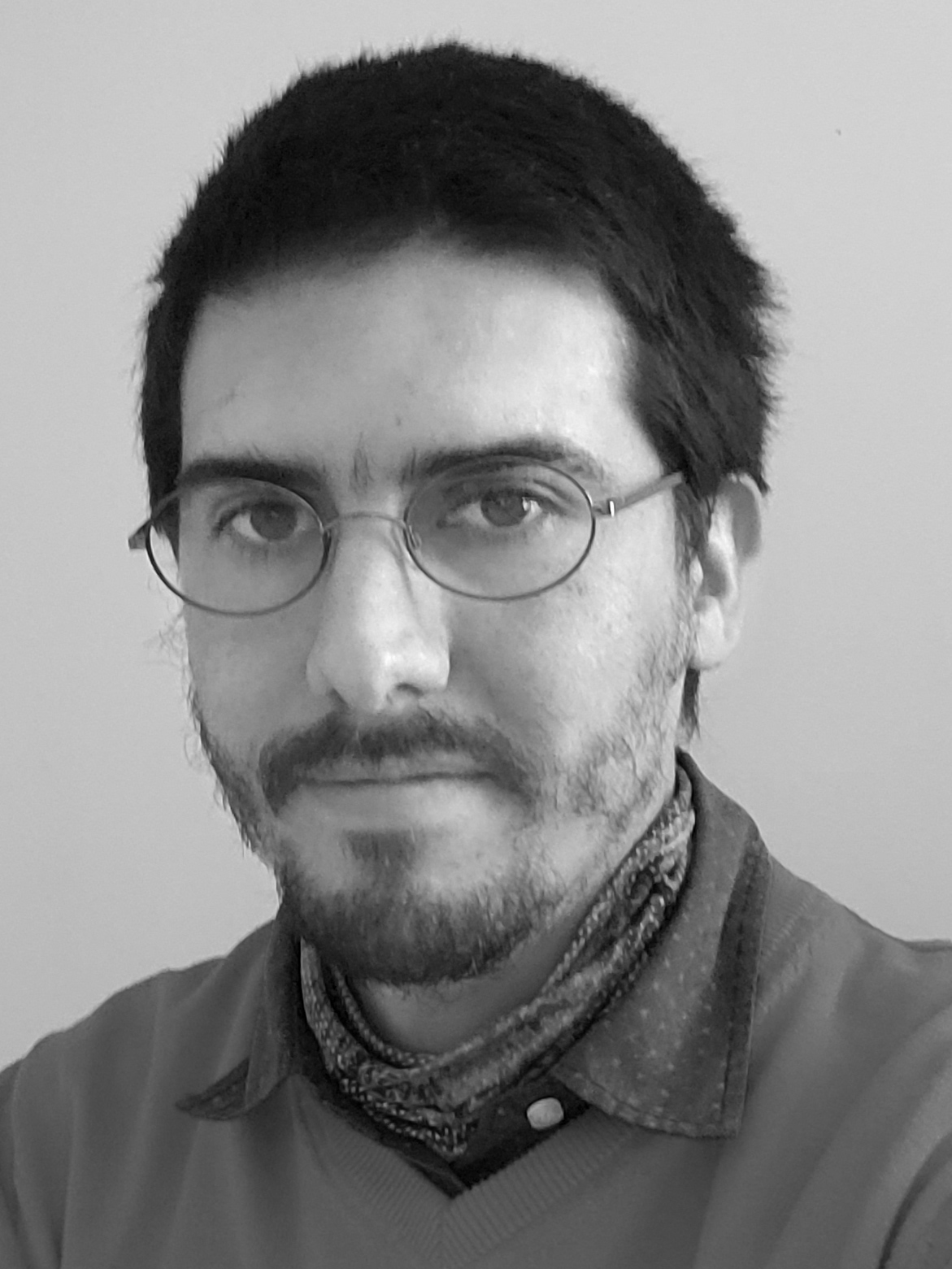}}]{Lorenzo Reyes-Chamorro}
	(S'13-M'16-SM'20) was born in Santiago, Chile in 1984.
    He received the B.Sc. degree in electrical engineering from the University of Chile, Santiago, Chile in 2009 and the Ph.D. degree at the Swiss Federal Institute of Technology of Lausanne (EPFL), Lausanne, Switzerland in 2016.
    He was a postdoctoral fellow at the Distributed Electrical System Laboratory, EPFL from 2016 to 2018. He is currently Assistant Professor in the Institute of Electricity and Electronics and Director of the Innovative Energy Technologies Center (INVENT UACh) at Universidad Austral de Chile, Valdivia, Chile.
\end{IEEEbiography}

\begin{IEEEbiography}[{\includegraphics[width=1in,height=1.25in,clip,keepaspectratio]{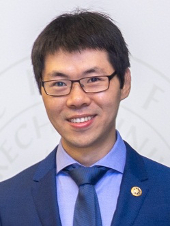}}]{Zhixiang Zou}
	(S’12-M’18-SM’20) received the B.Eng. and Ph.D. degrees in electrical and engineering from Southeast University, Nanjing, China, in 2007 and 2014, respectively, the Dr.-Ing. degree (summa cum laude) from Kiel University, Germany, in 2019. 
	He was an engineer in the State Grid Electric Power Research Institute, Nanjing, China, from 2007 to 2009. 
	He was a research fellow at the Chair of Power Electronics, Kiel University, Germany, from 2014 to 2019. 
	He is now an associate professor in the School of Electrical Engineering at the Southeast University. 
	His research interests include smart transformers, microgrid stability, modeling and control of power converters.
    Dr. Zou serves as an Associate Editor of the IEEE Open Journal of Power Electronics, an Associate Editor of the IEEE Access, an Editor of the International Transactions on Electrical Energy Systems, and an Editor of the Mathematical Problem in Engineering, and a Standing Director of IEEE PES Power System Relaying \& Control Satellite Committee.
\end{IEEEbiography}

\begin{IEEEbiography}[{\includegraphics[width=1in,height=1.25in,clip,keepaspectratio]{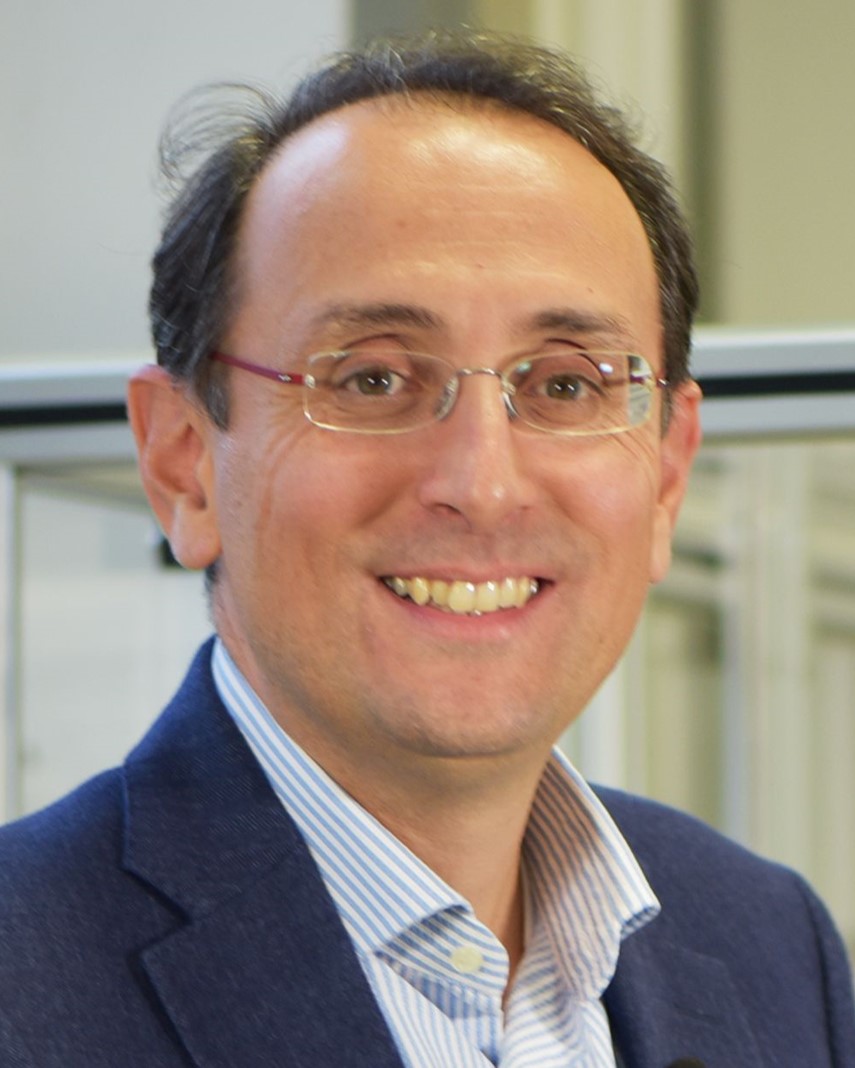}}]{Marco Liserre}
	(S'00-M'02-SM'07-F'13) received the MSc and PhD degree in Electrical Engineering from the Bari Polytechnic, respectively in 1998 and 2002. 
	He has been Associate Professor at Bari Polytechnic and from 2012 Professor in reliable power electronics at Aalborg University (Denmark). 
	From 2013 he is Full Professor and he holds the Chair of Power Electronics at Kiel University (Germany). 
	He has published 500 technical papers (1/3 of them in international peer-reviewed journals) and a book. These works have received more than 35000 citations. 
	Marco Liserre is listed in ISI Thomson report “The world’s most influential scientific minds” from 2014. 
    He has been awarded with an ERC Consolidator Grant for the project “The Highly Efficient And Reliable smart Transformer (HEART), a new Heart for the Electric Distribution System”.
    He is member of IAS, PELS, PES and IES. 
    He has been serving all these societies in different capacities. 
    He has received the IES 2009 Early Career Award, the IES 2011 Anthony J. Hornfeck Service Award, the 2014 Dr. Bimal Bose Energy Systems Award, the 2011 Industrial Electronics Magazine best paper award in 2011 and 2020 and the Third Prize paper award by the Industrial Power Converter Committee at ECCE 2012, 2012, 2017 IEEE PELS Sustainable Energy Systems Technical Achievement Award and the 2018 IEEE-IES Mittelmann Achievement Award.
\end{IEEEbiography}

\begin{IEEEbiography}[{\includegraphics[width=1in,height=1.25in,clip,keepaspectratio]{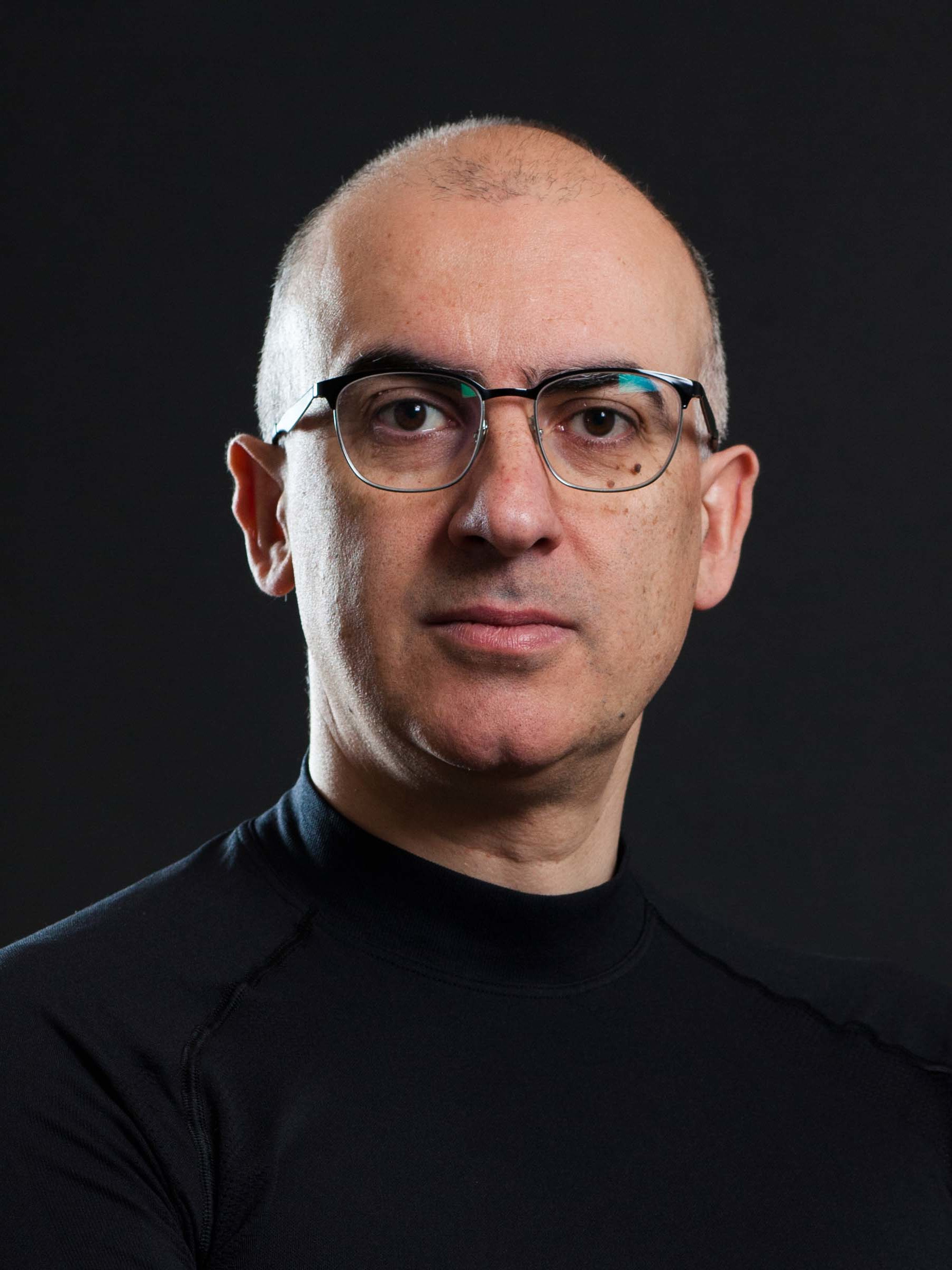}}]{Mario Paolone}
	(M’07–SM’10) received the M.Sc. (Hons.) and Ph.D. degrees in electrical engineering from the University of Bologna, Italy, in 1998 and 2002. 
	In 2005, he was an Assistant Professor in power systems with the University of Bologna, where he was with the Power Systems Laboratory until 2011. 
	Since 2011, he has been with the Swiss Federal Institute of Technology, Lausanne, Switzerland, where he is Full Professor and the Chair of the Distributed Electrical Systems Laboratory. 
	His research interests focus on power systems with reference to real-time monitoring and operational aspects, power system protections, dynamics and transients. 
	Dr. Paolone has authored or co-authored over 300 papers published in mainstream journals and international conferences in the area of energy and power systems that received numerous awards including the IEEE EMC Technical Achievement Award, two IEEE Transactions on EMC best paper awards, the IEEE Power System Dynamic Performance Committee’s prize paper award and the Basil Papadias best paper award at the 2013 IEEE PowerTech. Dr. Paolone was the founder Editor-in-Chief of the Elsevier journal Sustainable Energy, Grids and Networks.
\end{IEEEbiography}

\end{document}